\documentclass[usenatbib]{mn2e}
\usepackage{epsfig,lscape}

\newcommand{\etal}{et~al.} 
\newcommand{\ionhy}{H{\sc ii} }
\newcommand{\UCHII}{UCH{\sc ii} }

\newcommand{\kms}{$\mbox{km~s}^{-1}$ }
\newcommand{\kmsns}{$\mbox{km~s}^{-1}$}

\newcommand{\vsfig}[2]           
{
  \begin{center}
    \begin{minipage}[t]{0.05\textwidth}
      {\footnotesize \raisebox{40mm}{(#2)}}
    \end{minipage}
    \begin{minipage}[t]{0.42\textwidth}
      \psfig{file=./#1.ps,height=0.95\textwidth,angle=0}
    \end{minipage}
    \hfill
  \end{center}
}

\newcommand{\specdfig}[2]        
{
   \begin{center}
     \begin{minipage}[t]{0.45\textwidth}
         \psfig{file=eps/#1.eps,width=0.35\textwidth,width=0.8\textwidth,angle=270}
     \end{minipage}
     \hfill
     \begin{minipage}[t]{0.45\textwidth}
         \psfig{file=eps/#2.eps,width=0.35\textwidth,width=0.8\textwidth,angle=270}
     \end{minipage}
   \end{center}
}

\newcommand{\specsfig}[1]        
{
   \begin{center}
     \begin{minipage}[t]{0.45\textwidth}
         \psfig{file=eps/#1.eps,height=0.65\textwidth,width=1\textwidth,angle=0}
     \end{minipage}
   \end{center}
}

\newcommand{\boxfig}[1]        
{
   \begin{center}
     \begin{minipage}[t]{0.46\textwidth}
         \psfig{file=eps/#1.eps,height=0.45\textwidth,angle=0}
     \end{minipage}
   \end{center}
}

\newcommand{\twofig}[2]        
{
   \begin{center}
     \begin{minipage}[t]{0.5\textwidth}
         \psfig{file=eps/#1.eps,height=0.95\textwidth}
     \end{minipage}
     \hfill
     \begin{minipage}[t]{0.5\textwidth}
         \psfig{file=eps/#2.eps,height=0.95\textwidth}
     \end{minipage}
   \end{center}
}

\begin{document}

\title[12.2-GHz catalogue: longitudes 10$^{\circ}$ to 20$^{\circ}$]{12.2-GHz methanol maser MMB follow-up catalogue - III. Longitude range 10$^{\circ}$ to 20$^{\circ}$}
\author[S.\ L.\ Breen \etal]{S.\ L. Breen,$^{1}$\thanks{Email: Shari.Breen@csiro.au} S.\ P. Ellingsen,$^2$ J.\ L. Caswell,$^1$ J.\ A.\ Green,$^{1,3}$ M.\ A.\ Voronkov,$^1$ \newauthor A.\ Avison,$^{4,5}$ G.\ A.\ Fuller,$^{4,5}$  L.\ J.\ Quinn,$^4$ A.\ Titmarsh$^{2,1}$\\
 \\
  $^1$ CSIRO Astronomy and Space Science, Australia Telescope National Facility, PO Box 76, Epping, NSW 1710, Australia;\\
  $^2$ School of Mathematics and Physics, University of Tasmania, Private Bag 37, Hobart, Tasmania 7001, Australia;\\
  $^3$ SKA Organisation, Jodrell Bank Observatory, Lower Withington, Macclesfield, Cheshire SK11 9DL, UK;\\
  $^4$ Jodrell Bank Centre for Astrophysics, Alan Turing Building, School of Physics and Astronomy, University of Manchester,\\ Manchester M13 9PL, UK\\
  $^5$ UK ALMA Regional Centre node }
 \maketitle
  
 \begin{abstract}
We present the third instalment of a series of catalogues presenting 12.2-GHz methanol maser observations made towards each of the 6.7-GHz methanol masers detected in the Methanol Multibeam (MMB) Survey. The current portion of the catalogue includes the Galactic longitude region 10$^{\circ}$ to 20$^{\circ}$, where we detect 47 12.2-GHz methanol masers towards 99 6.7-GHz targets. We compare the occurrence of 12.2-GHz methanol masers with water maser emission, for which all 6.7-GHz methanol masers in the 6$^{\circ}$ to 20$^{\circ}$ longitude range have now been searched.  We suggest that the water masers follow a more complicated evolutionary scenario than has been found for the methanol and OH masers, likely due to their different pumping mechanisms. Comparisons of the 6.7-GHz methanol to OH maser peak flux density ratio and the luminosity of the associated 12.2-GHz sources suggests that the 12.2-GHz maser luminosity begins to decline around the time that an OH maser becomes detectable.


\end{abstract}


\section{Introduction}

Methanol masers at 6.7-GHz are an important tracer of young high-mass star formation regions, which they exclusively trace \citep[e.g.][]{Minier03,Xu08,Breen13}. They are readily detectable in these regions, with more than 1000 sources expected throughout the Galactic plane \citep[e.g.][]{CasMMB10,GreenMMB10,CasMMB102,Green12}. Slightly less common are the 12.2-GHz methanol masers, which are always found to have 6.7-GHz counterparts, and rarely surpass the flux density of the associated 6.7-GHz methanol maser emission \citep[e.g.][]{Caswell95b,Blas04,Gay,BreenMMB12a,BreenMMB12b}. The two transitions are typically found to be co-spatial to within a few milliarcseconds \citep[e.g.][]{Mos02} and can therefore be utilised in combination to probe the physical conditions in their immediate environments.

The conditions required to produce luminous methanol masers in the 6.7- and 12.2-GHz transitions are similar \citep[e.g.][]{Cragg05}, meaning that the presence or absence of a 12.2-GHz methanol maser may depend only on small changes in the properties of the environment. These changes occur alongside the evolution of the associated high-mass star formation region, and as such it has been proposed that the presence or absence of these masers (together with other common maser species) can be used to indicate the relative evolutionary stage of the associated star formation region \citep[e.g.][]{Ellingsen07,Breen10a}.

The MMB survey has searched for 6.7-GHz methanol masers in the Galactic plane from longitude 186$^{\circ}$ (through the Galactic centre) to 60$^{\circ}$ within latitudes of $\pm$2$^{\circ}$ \citep{Green09}. The initial survey was conducted with the Parkes 64-m radio telescope and all detected 6.7-GHz methanol maser sources were followed up with interferometric observations to gain precise positions. MMB source catalogues are currently available for the 186$^{\circ}$ (through the Galactic centre) to 20$^{\circ}$ longitude range \citep{CasMMB10,GreenMMB10,CasMMB102,Green12}, presenting a total of 707 sources.

We have targeted all 99 6.7-GHz methanol masers detected in the MMB survey in the Galactic longitude range 10$^{\circ}$ to 20$^{\circ}$. This catalogue forms the third instalment of our 12.2-GHz maser catalogue series, with the first catalogue presenting the longitude range 330$^{\circ}$ (through the Galactic Centre) to 10$^{\circ}$ \citep{BreenMMB12a}, and the second presenting the range 186$^{\circ}$ to 330$^{\circ}$ \citep{BreenMMB12b}. The current catalogue completes the 12.2-GHz follow-up of all MMB detections published to date. The final longitude range of 20$^{\circ}$ to 60$^{\circ}$ will be published following the publication of the MMB targets. Our combined catalogues comprise the largest, most complete search for 12.2-GHz methanol masers ever conducted. From this large repository of data we have uncovered significant insights into the characteristics of these sources \citep{Breen10a,Breen12stats,BreenMMB12a,BreenMMB12b}.

\section{Observations and data reduction} 

Using the Parkes 64-m radio telescope we have conducted 12.2-GHz follow-up observations towards all 6.7-GHz methanol masers detected in the MMB survey in the Galactic longitude range 10$^{\circ}$ to 20$^{\circ}$. Data were taken over three separate observing sessions; 2008 June 20-25, 2008 Dec 5-10 and 2010 March 19-23. An effort was made to try and minimise the time between the final MMB 6.7-GHz observations (chiefly made during 2008 March, August and 2009 March) and our follow-up 12.2-GHz observations (mostly made during the 2008 Dec session for sources in this longitude range). Observations were targeted towards precise positions derived from interferometric follow-ups of the 6.7-GHz methanol maser detections.

A detailed account of the 12.2-GHz observation strategy is presented in \citet{Breen12stats}, which we briefly summarise here. The observations were made with the {\em Ku}-band receiver, which detected two orthogonal linear polarisations, on the Parkes 64-m radio telescope. The system equivalent flux densities were 205 and 225~Jy for the respective polarisations during observations in 2008 June, and slightly higher at 220 and 240~Jy during 2008 December and 2010 March. At 12.2-GHz, the Parkes radio telescope has a half-power-beam-width of 1.9~arcmin and the telescope has rms pointing errors of $\sim$10~arcsec. The Parkes multibeam correlator was configured to record 8192 channels across a 16~MHz bandwidth. This configuration resulted in a usable velocity coverage of $\sim$290~\kms and a velocity resolution of 0.08~\kms after Hanning smoothing (applied during the data processing). Flux density calibration was achieved from daily observations of PKS~B1934--638, which, at 12.2 GHz has a flux density of 1.825~Jy \citep{Sault03}, and was stable over each of the observing sessions. We estimate the uncertainty in our flux density calibration to be $\sim$10 per cent.

Each of the target 6.7-GHz methanol masers was observed for at least 5 minutes and up to 20 minutes over the three observing epochs. Targets that were observed during more than one epoch were first inspected individually for maser emission and then averaged together to identify additional weak sources. Whenever possible, 12.2-GHz non-detections were observed on more than one epoch to try and minimise the number of non-detections due to variability. Data were reduced using the ATNF (Australia Telescope National Facility) Spectral Analysis Package ({\sc asap}). 
Velocities are with respect to LSR, with an adopted rest frequency of 12.178597 GHz \citep{Muller04} and alignment of velocity channels was carried out during processing. 
The resultant 5-$\sigma$ detection limits predominantly lie in the range 0.55 to 0.80~Jy for a single epoch, and are comparable to the 5-$\sigma$ detection limit of 0.85~Jy achieved in the MMB survey \citep{Green09}. 

All observations were made at a fixed frequency of 12~178 MHz allowing us to create a sensitive bandpass using the median value for each spectral channel from all the observations made over each observing run. This strategy resulted in a smaller noise contribution to the quotient spectrum than can be achieved by the usual method of obtaining a unique reference spectrum for each source. However, this strategy introduced a baseline ripple which we removed by subtracting a running median over 100 channels from each final spectrum; an effective solution for narrow lines such as masers, but one that makes it difficult to recover broader features such as those created by absorption since they can be comparable to the ripple in both width and amplitude. A further artefact (discussed in \citealt{BreenMMB12a}) can arise when very strong maser emission (100's of Jy)  is present over large velocity ranges and results in small negative dips in amplitude either side of the strong maser emission, but it is unlikely that any spectra in the 10$^{\circ}$ to 20$^{\circ}$ longitude range have been affected in this way given that the strongest source had a peak flux density of 29~Jy.

\section{Results}\label{sect:results}

We have detected 47 12.2-GHz methanol masers towards the 99 6.7-GHz methanol maser targets in the 10$^{\circ}$ to 20$^{\circ}$ longitude range \citep{GreenMMB10}. This equates to a detection rate of 47 per cent over this longitude range, and combining with the detection statistics of the longitude 186$^{\circ}$ (through the Galactic centre) to 10$^{\circ}$ \citep{BreenMMB12a,BreenMMB12b} results in an overall detection rate of 45 per cent towards the 707 6.7-GHz methanol masers observed in this series of follow-up catalogues. We attribute the small variation in the detection statistics to a combination of locality (\citealt{Breen12stats} found there to be real variations in the detection statistics in different regions of the Galactic plane), individual source variability and our somewhat uneven detection limit (i.e. some sources had longer integration times than others). We estimate the detection rate of 12.2-GHz methanol masers towards complete samples of 6.7-GHz methanol masers to fall within $\pm$5 per cent of our overall detection rate of 45 per cent (for searches with similar sensitivity).

The 6.7-GHz methanol maser targets are listed in Table~\ref{tab:6MMB}, which includes the characteristics of the emission as detected in the MMB survey `MX' spectra. The integrated flux densities of both the 6.7- and 12.2-GHz methanol have proven to be useful quantities for comparison \citep[e.g.][]{Breen12stats}. Since this value is not given in the MMB catalogues for the 6.7-GHz methanol maser sources, we have determined it using the MMB `MX' spectra, and for internal consistency, we also derived all other source properties independently. We find there to be only minor discrepancies between the values reported in \citet{GreenMMB10} and those listed here and these differences have mostly arisen due to the usage of data from slightly different epochs. 

Table~\ref{tab:6MMB} also gives either the 12.2-GHz 5-$\sigma$ detection limits at each epoch, or an indication that a 12.2-GHz source was detected. In the case where 12.2-GHz observations were completed over two or more epochs, the data-averaged 5-${\sigma}$ detection limits are generally a factor of $\sqrt{2}$ better than those listed. We were able to reliably identify some masers at the 3- or 4-$\sigma$ limit when there was excellent velocity correspondence with a 6.7-GHz methanol maser feature. Sources requiring further detailed comments to fully convey their nature are marked with an `*' in Table~\ref{tab:6MMB} and discussed individually in Section~\ref{sect:ind}.

A similar MMB follow-up survey has been conducted for the 22-GHz water maser line and the first catalogue covering the 6$^{\circ}$ to 20$^{\circ}$ longitude range will soon be published \citep{Titmarsh13}. We have therefore been able to distinguish those 6.7-GHz methanol masers with and without associated water masers in Table~\ref{tab:6MMB}, and discuss the detectability of water masers compared to 12.2-GHz masers towards a complete sample of 6.7-GHz methanol masers for the first time in Section~\ref{sect:water}.

\begin{table*}\footnotesize
 \caption{Characteristics of the 6.7-GHz methanol maser targets as well as a brief description of the 12.2-GHz results (either detection or 5-$\sigma$ detection limits). The full complement of 12.2-GHz source properties are listed in Table~\ref{tab:12MMB}. Column 1 gives the Galactic longitude and latitude of each source and is used as an identifier (an `*' indicates sources with notes in Section~\ref{sect:ind}); column 2 gives the peak flux density (Jy) of the 6.7-GHz sources, derived from follow-up MX observations at the accurate 6.7-GHz position unless otherwise noted; columns 3 and 4 give the peak velocity and the minimum and maximum velocity (\kmsns) of the 6.7-GHz emission respectively (also derived from Parkes MX observations); column 5 gives the integrated flux density of the 6.7-GHz emission (Jy \kmsns). Column 6 shows those sources that are also associated with water maser emission \citep{Titmarsh13}. A `--' in either of the detection limit columns (i.e. columns 7, 9 or 11) indicates that no observations were made on the given epoch. The values listed in columns 7-12 are replaced with the word `detection' where 12.2-GHz emission is observed. The presence of `conf' in the columns showing the integrated flux density indicates that a value for this characteristic could not be extracted do to confusion from nearby sources. Columns 7-12 give the 5-$\sigma$ 12.2-GHz methanol maser detection limits and observed velocity ranges for the 2008 June, 2008 December and 2010 March epochs respectively.}
  \begin{tabular}{lllllclllclllcllllll} \hline
 \multicolumn{1}{c}{\bf Methanol maser} &\multicolumn{4}{c}{\bf 6.7-GHz properties} & {\bf  Water?} & \multicolumn{6}{c}{\bf 12.2-GHz observations}\\
    \multicolumn{1}{c}{\bf ($l,b$)}& {\bf S$_{6.7}$} & {\bf Vp$_{6.7}$} & {\bf Vr$_{6.7}$ } & {\bf  I$_{6.7}$} && \multicolumn{2}{c}{\bf 2008 June} & \multicolumn{2}{c}{\bf 2008 December} & \multicolumn{2}{c}{\bf 2010 March}\\	
      \multicolumn{1}{c}{\bf (degrees)}  &{\bf (Jy)} &&{\bf (\kmsns)}& & &\multicolumn{1}{c}{\bf 5-$\sigma$} & \multicolumn{1}{c}{\bf Vr} & \multicolumn{1}{c}{\bf 5-$\sigma$} & \multicolumn{1}{c}{\bf Vr} & \multicolumn{1}{c}{\bf 5-$\sigma$} & \multicolumn{1}{c}{\bf Vr}\\  \hline \hline
G\,10.205--0.345	&	2.0	&	6.6		&	6.3,10.4		&	1.1	  	&&	--		&				&	$<$0.60	&	--200,140		&	$<$0.78	&	--140,200	\\
G\,10.287--0.125 	&	8.1	& 	8.0		& 	1.7,8.3		&	11	  	&y&	$<$0.50	&	--140,200		&	$<$0.80	&	--200,140		&	$<$1.2	&	--140,200\\
G\,10.299--0.146 	&	0.9	&	19.9		&	19.3,20.3		&	0.7	  	&&	$<$0.55	&	--140,200		&	$<$0.80	&	--200,140		&	$<$0.85	&	--140,200	\\
G\,10.320--0.259 	&	9.5	&	39.0		&	25.6,39.9		&	13		&y&	 \multicolumn{6}{c}{detection}  \\
G\,10.323--0.160* 	&	90	&	11.5		&	2.4,17.8		&	148	 	&&	 \multicolumn{6}{c}{detection} \\
G\,10.342--0.142* 	&	15	&	15.4		&	7.0,17.0		&	22	  	&y&	 \multicolumn{6}{c}{detection}\\
G\,10.356--0.148 	&	0.8	&	50.0		&	49.6,50.9		&	0.3	 	&&	--	&			&	--	&			&	$<$0.8	&	--140,200	 \\
G\,10.444--0.018 	&	24	&	73.3		&	67.6,79.0		&	conf	  	&y&	\multicolumn{6}{c}{detection}\\
G\,10.472+0.027*	& 	33	&	75.1	  	&	57.5,77.6		&	conf		&y&	\multicolumn{6}{c}{detection} \\
G\,10.480+0.033 	&	23	&	59.5		&	57.0,66.0		&	conf  	&&	$<$0.70	&	--140,200		&	--	&	&	--\\
G\,10.627--0.384* 	&	3.9	&	4.6		&	--6.0,7.1		&	3.3		&&       \multicolumn{6}{c}{detection}  \\
G\,10.629--0.333 	&	6.1	&	--8.0		&	--13.0,--1.0	&	11	  	&&	$<$0.53	&	--140,200		&	$<$0.80	&	--190,150	&	--\\
G\,10.724--0.334 	&	4.8	&	--2.2		&	--2.5,--1.7		&	1.9		&y&	$<$0.70	&	--140,200		&	$<$0.72	&	--190,150	&	--  \\
G\,10.822--0.103 	&	0.3	&	69.1		&	68.8,69.4		&	0.1	 	&y&	$<$0.72	&	--140,200		&	$<$0.75	&	--190,150 		&	$<$0.75	&	--140,200\\
G\,10.886+0.123* 	&	12	&	17.1		&	14.5,22.0		&	14	  	&y&	\multicolumn{6}{c}{detection} \\
G\,10.958+0.022 	&	12	&	24.5		&	23.7,25.1		&	9.5		&y&       \multicolumn{6}{c}{detection}  \\
G\,11.034+0.062 	&	0.5	&	20.5		&	15.8,20.8		&	0.2	  	&y &	$<$0.74	&	--140,200		&	$<$0.77	&	--190,150		&	--\\
G\,11.109--0.114 	&	15	&	32.3		&	22.8,34.1		&	25	 	&&	 \multicolumn{6}{c}{detection} \\
G\,11.497--1.485 	&	68	&	6.6		&	4.5,17.1		&	153	  	&&	 \multicolumn{6}{c}{detection} \\
G\,11.903--0.102 	&	11	&	33.8		&	33.4,36.8		&	7.3	  	&&	 \multicolumn{6}{c}{detection} 	\\
G\,11.904--0.141 	&	65	&	42.8		&	39.9,44.2		&	65	  	&&	 \multicolumn{6}{c}{detection} \\
G\,11.936--0.150 	&	2.2	&	48.5		&	45.9,49.0		&	3.2  		&&	$<$0.74	&	--140,200		&	$<$0.80	&	--190,150		&	$<$0.54	&	--140,200\\
G\,11.936--0.616* 	&	43	&	32.3		&	30.1,44.3		&	63  		&&	\multicolumn{6}{c}{detection}	\\
G\,11.992--0.272 	&	1.9	&	59.7		&	56.3,61.3		&	1.5	 	&&	$<$0.74	&	--140,200		&	$<$0.75	&	--190,150		&	--\\
G\,12.025--0.031 	&	96	&	108.3	&	104.9,112.8	&	114		&&	\multicolumn{6}{c}{detection} \\
G\,12.112--0.126 	&	3.0	&	39.9		&	38.4,50.6		&	3.1		&&	\multicolumn{6}{c}{detection}   \\
G\,12.181--0.123 	&	1.9	&	29.6		&	29.3,31.0		&	1.7		&&	$<$0.75	&	--140,200	  	&	$<$0.78	&	--190,150		&	--\\
G\,12.199--0.033 	&	14	&	49.3		&	48.3,57.1		&	17		&y&	$<$0.74	&	--140,200		&	$<$0.79	&	--190,150	 	&	-- \\
G\,12.202--0.120 	&	0.7	&	26.3		&	26.3,26.7		&	0.3	  	&&	$<$0.73	&	--140,200		&	$<$0.79	&	--190,150		&	-- \\
G\,12.203--0.107 	&	2.4	&	20.5		&	20.0,32.0		&	conf		&&	$<$0.74	&	--140,200		&	$<$0.58	&	--190,150 		&	$<$0.79	&	--140,200\\
G\,12.209--0.102 	&	11	&	19.8		&	16.0,22.0		&	conf		&y&  \multicolumn{6}{c}{detection}\\
G\,12.265--0.051* 	&	2.2	&	68.5		&	60.2,70.6		&	5.3		&y&	 \multicolumn{6}{c}{detection}\\
G\,12.526+0.016 	&	3.1	&	42.6		&	39.0,43.6		&	3.7		&&	\multicolumn{6}{c}{detection}  \\
G\,12.625--0.017 	&	26	&	21.6		&	21.1,28.1		&	26	  	&&	\multicolumn{6}{c}{detection}		\\
G\,12.681--0.182 	&	351	&	57.5		&	50.3,61.6		&	836	  	&y& \multicolumn{6}{c}{detection} \\
G\,12.776+0.128 	&	0.5	&	33.0		&	30.8,33.1		&	0.3	 	&&	$<$0.74	&	--140,200 		&	$<$0.80	&	--190,150		&	$<$0.78	&	--140,200	\\
G\,12.889+0.489 	&	78	&	39.3		&	27.8,43.0		&	100	 	&y& \multicolumn{6}{c}{detection} \\
G\,12.904--0.031 	&	27	&	59.0		&	57.7,60.6		&	28		&y&	\multicolumn{6}{c}{detection}    \\
G\,12.909--0.260 	&	251	&	39.9		&	35.3,47.0		&	303		&y& \multicolumn{6}{c}{detection}   \\
G\,13.179+0.061 	&	0.8	&	46.6		&	45.9,50.2		&	0.5	  	&&	$<$0.73	&	--140,200		&	$<$0.80	&	--190,150		&	--\\
G\,13.657--0.599 	&	32	&	51.2		&	45.2,52.3		&	40	 	&y&	$<$0.74	&	--140,200	 	&	$<$0.77	&	--190,150		&	--\\
G\,13.696--0.156 	&	1.9	&	99.3		&	98.5,108.3	&	2.8	  	&& \multicolumn{6}{c}{detection} \\
G\,13.713--0.083 	&	13	&	43.6		&	42.2,52.9		&	9.0	  	&& 	$<$0.76		&	--140,200		&	$<$0.57		&	--190,150\\
G\,14.101+0.087 	&	86	&	15.3		&	4.2,17.9		&	141	 	&y& \multicolumn{6}{c}{detection}	 \\
G\,14.230--0.509 	&	3.6$^{sc}$& 25.3	&	24.6,26.7		&	1.3	  	&&	--		&				&		--	&			&	$<$0.79	&	--140,200\\
G\,14.390--0.020 	&	3.1	&	26.9		&	25.0,28.3		&	2.5		&&\multicolumn{6}{c}{detection}  \\
G\,14.457--0.143* 	&	0.8	&	43.1		&	38.0,43.6		&	0.7		&&	--	&		&	$<$1.2		&	--190,150		&	--\\
G\,14.490+0.014 	&	1.3	&	20.2		&	20.1,24.2		&	0.5	 	&&	--	&		&	$<$0.57		&	--190,150		&	-- \\
G\,14.521+0.155 	&	1.6	&	3.4		&	--2.2,5.8		&	2.4	  	&& \multicolumn{6}{c}{detection}\\
G\,14.604+0.017 	&	2.3	&	24.6		&	22.2,35.1		&	2.9	  	&y&	\multicolumn{6}{c}{detection} \\
\end{tabular}\label{tab:6MMB}
\end{table*}

\begin{table*}\addtocounter{table}{-1}
  \caption{-- {\emph {continued}}}
  \begin{tabular}{lllllclllclllcll} \hline
 \multicolumn{1}{c}{\bf Methanol maser} &\multicolumn{4}{c}{\bf 6.7-GHz properties} &{\bf Water?}& \multicolumn{6}{c}{\bf 12.2-GHz observations}\\
    \multicolumn{1}{c}{\bf ($l,b$)}& {\bf S$_{6.7}$} & {\bf Vp$_{6.7}$} & {\bf Vr$_{6.7}$ } & {\bf  I$_{6.7}$} && \multicolumn{2}{c}{\bf 2008 June} & \multicolumn{2}{c}{\bf 2008 December} & \multicolumn{2}{c}{\bf 2010 March}\\	
      \multicolumn{1}{c}{\bf (degrees)}  &{\bf (Jy)} && {\bf (\kmsns)}& && \multicolumn{1}{c}{\bf 5-$\sigma$} & \multicolumn{1}{c}{\bf Vr} & \multicolumn{1}{c}{\bf 5-$\sigma$} & \multicolumn{1}{c}{\bf Vr} & \multicolumn{1}{c}{\bf 5-$\sigma$} & \multicolumn{1}{c}{\bf Vr}\\  \hline \hline
G\,14.631--0.577 	&	1.1	&	25.2		&	24.7,25.5		&	0.5		& &	--	&		&	$<$0.59	&	--190,150	 &	--\\
G\,14.991--0.121 	&	7.3	&	46.0		&	44.8,53.9		&	3.5		 &&	--	&		&	$<$0.57	&	--190,150		&	--	\\
G\,15.034--0.677 	&	47	&	21.3		&	20.1,23.8		&	17		&&	\multicolumn{6}{c}{detection}   \\
G\,15.094+0.192 	&	14	&	25.7		&	22.5,26.2		&	8.0	 	&y&	\multicolumn{6}{c}{detection} \\
G\,15.607--0.255 	&	0.8	&	66.0		&	65.6,66.1		&	0.2	  	&&	--	&		&	$<$0.57	&	--190,150		&	$<$0.80	&	--140,200	\\
G\,15.665--0.499 	&	43	&	--2.9		&	--4.5,--2.4		&	20		 &y&	--	&		&	$<$0.57	&	--190,150 		&	--	\\
G\,16.112--0.303 	&	2.2	&	34.4		&	34.2,34.9		&	1.0	  	&&	--	&		&	$<$0.58	&	--190,150		&	--	\\
G\,16.302--0.196 	&	11	&	51.9		&	41.1,52.7		&	12	  	&& \multicolumn{6}{c}{detection}\\
G\,16.403--0.181 	&	0.6	&	39.1		&	39.0,39.3		&	0.2	  	&&	--	&		&	$<$0.57	&	--190,150		&	--\\
G\,16.585--0.051 	&	37	&	62.1		&	56.4,68.8		&	50		&y&	\multicolumn{6}{c}{detection}   \\
G\,16.662--0.331 	&	3.2	&	43.0		&	42.5,43.8		&	1.5	 	&&	--	&		&	$<$0.67	&	--190,150	&	--	 \\
G\,16.831+0.079 	&	4.1	&	58.7		&	58.4,68.8		&	7.4	  	&y&	--	&		&	$<$0.56	&	--190,150	&	--	\\
G\,16.855+0.641 	&	1.7	&	24.2		&	23.3,24.9		&	0.9	  	&&	--	&		&	$<$0.56	&	--190,150	&	--	\\
G\,16.864--2.159 	&	29	&	15.0		&	14.6,19.8		&	20	  	&y&	$<$0.52	&	--140,200	&	$<$0.80	&	--190,150	&	$<$0.55	&	--140,200	\\
G\,16.976--0.005 	&	0.7	&	6.6		&	5.7,8.5		&	0.8	  	&&	\multicolumn{6}{c}{detection}\\
G\,17.021--2.403 	&	4.9	&	23.6		&	17.4,24.6		&	7.2	  	&y&	--	&		&	$<$0.57	&	--190,150	&	--	\\
G\,17.029--0.071 	&	1.1	&	91.4		&	90.9,95.7		&	0.9	  	&&	--	&		&	$<$0.58	&	--190,150	&	--	\\
G\,17.638+0.157 	&	25	&	20.8		&	20.5,21.0		&	5.9	 	&y&	--	&		&	$<$0.57	&	--190,150 	&		--\\
G\,17.862+0.074 	&	1.4	&	110.6	&	107.9,120.2	&	3.2		& &	\multicolumn{6}{c}{detection}  \\
G\,18.073+0.077 	&	5.7	&	55.6		&	44.0,57.4		&	9.7	 	&&	\multicolumn{6}{c}{detection} \\
G\,18.159+0.094 	&	8.5	&	58.4		&	47.5,59.3		&	6.4	  	&&	--	&		&	$<$0.56	&	--190,150 &	--\\
G\,18.262--0.244 	&	23	&	74.3		&	72.9,80.8		&	40	  	&&	\multicolumn{6}{c}{detection}	\\
G\,18.341+1.768 	&	96	&	28.0		&	27.0,32.0		&	49	  	&y&	\multicolumn{6}{c}{detection}	\\
G\,18.440+0.045 	&	1.9	&	61.8		&	57.9,65.5		&	3.1		&&	--	&		&	$<$0.56	&	--190,150	 &	-- \\
G\,18.460--0.004 	&	24	&	49.4		&	47.1,49.7		&	15	  	&&	--	&		&	$<$0.57	&	--190,150	&	--\\
G\,18.661+0.034* 	&	8.9	&	79.0		&	76.0,83.0		&	conf	  	&y&	--	&		&	$<$0.58	&	--190,150	&$<$0.50	&	--140,200\\
G\,18.667+0.025* 	&	8.7	& 	76.6		&	76.1,81.0		&	conf	  	&&	\multicolumn{6}{c}{detection}	\\
G\,18.733--0.224 	&	2.3	&	45.8		&	44.5,48.9		&	4.2	  	&&	--	&		&	--	&		&	$<$0.54	&	--140,200	\\
G\,18.735--0.227 	&	4.7	&	37.9		&	37.6,38.6		&	2.3	  	&y&	--	&		&	--	&		&	$<$0.54	&	--140,200\\
G\,18.834--0.300 	&	5.0	&	41.3		&	38.7,43.0		&	3.9	 	&&	--	&		&	$<$0.56	&	--190,150	 &	--\\
G\,18.874+0.053 	&	14	&	38.7		&	37.9,39.6		&	6.8	  	&&	\multicolumn{6}{c}{detection}	\\
G\,18.888--0.475 	&	5.7	&	56.5		&	52.3,57.7		&	7.1	  	&&	--	&		&	$<$0.56	&	--190,150		&	--\\
G\,18.999--0.239 	&	0.7	&	69.5		&	69.2,69.6		&	0.2	  	&y&	--	&		&	--		&			&	$<$0.54	&	--140,200\\
G\,19.009--0.029*	&	19	&	55.3		&	53.8,60.9		&	29	  	&y&	\multicolumn{6}{c}{detection}\\
G\,19.249+0.267* 	&	2.6	&	20.5		&	12.6,24.8		&	4.0	  	&&	\multicolumn{6}{c}{detection}\\
G\,19.267+0.349 	&	5.3	&	16.2		&	13.3,17.0		&	3.3	  	&y&	--	&		&	$<$0.56	&	--190,150	&	--	\\
G\,19.365--0.030 	&	34	&	25.2		&	24.4,30.0		&	22	  	&&	\multicolumn{6}{c}{detection}\\
G\,19.472+0.170n 	&	18	&	21.7		&	17.0,23.0		&	conf	  	&y&	\multicolumn{6}{c}{detection} \\
G\,19.472+0.170 	&	3.3	&	13.8		&	12.7,17.7		&	conf	  	&&	--	&		&	$<$0.80	&	--190,150		&	$<$0.55	&	--140,200\\
G\,19.486+0.151 	&	6.0	&	20.6		&	19.0,27.5		&	conf	  	&y&	\multicolumn{6}{c}{detection} \\
G\,19.496+0.115 	&	7.5	&	121.2	&	120.6,121.7	&	3.3	  	&y&	--	&		&	$<$0.56	&	--190,150	&	--	\\
G\,19.609--0.234 	&	0.7	&	40.2		&	38.9,41.8		&	0.5	  &y&	$<$0.52	&	--140,200		&	$<$0.80	&	--190,150	&	--	\\
G\,19.612--0.120 	&	0.9	&	53.1		&	52.6,53.5		&	0.5	  &&	$<$0.52	&	--140,200		&	--	&		&	-- \\
G\,19.612--0.134 	&	13	&	56.5		&	49.3,60.5		&	13	  &y&	\multicolumn{6}{c}{detection}  \\
G\,19.614+0.011 	&	4.0	&	32.8		&	30.9,34.8		&	5.7	  &y&	\multicolumn{6}{c}{detection}\\
G\,19.667+0.114* 	&	2.2	&	14.2		&	13.9,17.6		&	2.9	  &&	\multicolumn{6}{c}{detection}	\\
G\,19.701--0.267 	&	11	&	43.9		&	42.0,46.2		&	13	  &y&	--	&		&	$<$0.80	&	--190,150& --\\
G\,19.755--0.128 	&	3.6	&	123.1	&	116.2,123.4	&	1.9	  &&	--	&		&	$<$0.80	&	--190,150	& --\\
G\,19.884--0.534 	&	4.7	&	46.8		&	46.1,47.8		&	3.0	  &y&	--	&		&	$<$0.80	&	--190,150	&--	\\ \hline
\end{tabular}
\end{table*}

The characteristics of the detected 12.2-GHz methanol masers are given in Table~\ref{tab:12MMB} and list the position of the 6.7-GHz methanol maser target, followed by the 12.2-GHz peak flux density, velocity of the peak 12.2-GHz emission, velocity range and integrated flux density. Following some of the source names, references are given to previous 12.2-GHz maser detections. Our search of the literature showed that 30 of the 47 (64 per cent) 12.2-GHz methanol maser sources we detect are new to our search.

\begin{table*}
\caption{Characteristics of the detected 12.2-GHz methanol masers. Column 1 gives the Galactic longitude and latitude of each source and is used as an identifier; columns 2 and 3 give the equatorial coordinates for each of the 6.7-GHz methanol masers, derived from interferometric observations \citep{GreenMMB10}. References for previously detected 12.2-GHz sources follow the source name and are as follows; 1: \citet{Breen10a}; 2: \citet{Caswell95b}; 3: ~\citet{Gay}; 4: \citet{Koo88}; 5: \citet{Catarzi93}; 6: \citet{Kemball88}; 7: \citet{Cas93}; 8: \citet{Norris1987}; 9: \citet{Blas04}; 10: \citet{Batrla87}, and are presented as superscripts. Column 4 gives the epoch of the observed data (sometimes indicating that the presented data is the average of multiple epochs) presented in columns 5 - 8 which give the peak flux density (Jy) or 5-$\sigma$ detection limit, velocity of the 12.2-GHz peak (\kmsns), velocity range (\kmsns) and integrated flux density (Jy \kmsns), respectively.} 
  \begin{tabular}{lllllccclclllclllcll} \hline
\multicolumn{1}{c}{\bf Methanol maser} &\multicolumn{1}{c}{\bf RA} & \multicolumn{1}{c}{\bf Dec}  &{\bf Epoch}&  {\bf S$_{12.2}$} & {\bf Vp$_{12.2}$} & {\bf Vr$_{12.2}$} & \multicolumn{1}{c}{\bf I$_{12.2}$} \\
    \multicolumn{1}{c}{\bf ($l,b$)}&  \multicolumn{1}{c}{\bf (J2000)} & \multicolumn{1}{c}{\bf (J2000)}   & & {\bf (Jy)} & {\bf (\kmsns)} & {\bf  (\kmsns)} & \multicolumn{1}{c}{\bf (Jy \kmsns)}\\
    \multicolumn{1}{c}{\bf (degrees)}  & \multicolumn{1}{c}{\bf (h m s)}&\multicolumn{1}{c}{\bf ($^{o}$ $'$ $''$)}\\  \hline \hline
G\,10.320--0.259 	&	18 09 23.30 	&	--20 08 06.9 	&	  2008 Jun	&	3.9	&	36.3	&	35.9,39.2	&1.9\\
				&				&				&	  2010 Mar	&	3.7	&	36.3	&	35.9,39.2	&	1.9\\
G\,10.323--0.160$^{1}$ 	&	18 09 01.46 	&	--20 05 07.8 	&	  2008 Jun	&	2.3	&	10.2	&	7.4,12.4	&	1.4 \\
				&				&				&	  2010 Mar	&	2.8	&	10.2	&	9.8,11.2	&	1.6		\\
G\,10.342--0.142  	&	18 08 59.99 	&	--20 03 35.4 	&	  2008 Dec	&	5.2	&	7.5	&	7.1,8.6	&	2.7\\
				&				&				&	  2010 Mar	&	5.6	&	10.2	&	6.1,14.0	&	5.4	\\
G\,10.444--0.018$^{1,2}$  	&	18 08 44.88 	&	--19 54 38.2 	&	  2008 Jun	&	6.3	&	71.9	&	67.9,73.6	&	9.6\\
G\,10.472+0.027$^{1,2,4,5,6,7}$  	&	18 08 38.20 	&	--19 51 50.1 	&	  2008 Jun	&	12	&	75.0	&	73.6,76.7	&	8.9\\
				&				&				&	  2008 Dec	&	10	&	74.9	&	73.6,76.7	&	7.6	\\
G\,10.627--0.384$^{1,2,10}$  	&	18 10 29.22 	&	--19 55 41.1 	&	  2008 Jun	&	1.4	&	4.5	&	4.2,6.5	&	0.7	\\
G\,10.886+0.123 	&	18 09 07.98 	&	--19 27 21.8 	&	  2008 Jun	&	$<$0.72	&	&	--140,200		\\
				&				&				&	  2008 Dec	&	0.6	&	19.7	&	19.6,20.2	&	0.2		\\
G\,10.958+0.022 	&	18 09 39.32 	&	--19 26 28.0 	&	  2008 Jun	&	1.0 	& 24.5 	& 	23.9,25.0	&	0.6	\\
G\,11.109--0.114 	&	18 10 28.25 	&	--19 22 29.1 	&	  2008 Dec	&	5.0 	&	23.9	&	23.5,33.5	&	3.9 \\
G\,11.497--1.485$^{1,3}$  	&	18 16 22.13 	&	--19 41 27.1 	&	  2008 Jun	&	29	&	9.0	&	5.9,17.1	&	16	\\
G\,11.903--0.102 	&	18 12 02.70 	&	--18 40 24.7 	&	  2008 Jun	&	1.8 	&	33.8	&	33.6,35.8	&	0.5	\\
G\,11.904--0.141$^{1,2,7}$  	&	18 12 11.44 	&	--18 41 28.6 	&	  2008 Jun	&	17	& 43.0 	&	42.1,44.0	&	8.1	\\
G\,11.936--0.616$^{1,2}$  	&	18 14 00.89 	&	--18 53 26.6 	&	  2008 Jun	&	4.3 	&	39.8	&	31.5,42.2	&	4.5\\
G\,12.025--0.031$^{1,2,7}$  	&	18 12 01.86 	&	--18 31 55.7 	&	  2008 Jun	&	14	&	108.6	&	107.7,109.4	&	8.7	\\
G\,12.112--0.126 	&	18 12 33.39 	&	--18 30 07.6 	&	  2008 Jun	&	1.3 	&	40.3	&	40.1,40.6	&	0.4\\
G\,12.209--0.102 	&	18 12 39.92 	&	--18 24 17.9 	&	  2008 Dec	&	0.5	&	17.4 	&	17.3,18.4	&	0.2\\
				&				&				&	  2010 Mar	&	$<$0.79	&		&	--140,200\\
G\,12.265--0.051 	&	18 12 35.40 	&	--18 19 52.3 	&	  2008 Jun	&	$<$0.72	&	&	--140,200	 \\
				&				&				&	  2008 Dec	&	0.4	&68.3 &60.5,68.3 &0.1\\
G\,12.526+0.016 	&	18 12 52.04 	&	--18 04 13.6 	&	  2008 Jun	&	0.4 	&	42.5	&	42.2,	42.8 & 0.1	\\
				&				&				&	  2008 Dec	&	0.5 	&	42.3 	&42.1,43.0 	&0.2\\
G\,12.625--0.017 	&	18 13 11.30 	&	--17 59 57.6 	&	  2008 Dec	&	5.5 	& 21.6	&21.3,24.7	& 2.2\\
G\,12.681--0.182$^{1,2,4,5,7,8}$  	&	18 13 54.75 	&	--18 01 46.6 	&	  2008 Jun	&	8.0 	& 57.1	&55.6,60.3 	&	7.3	\\
				&				&				&	  2008 Dec	&	6.3 	& 57.1	&56.3,58.5 	&6.1 \\
G\,12.889+0.489$^{1,2,5,6}$  	&	18 11 51.40 	&	--17 31 29.6 	&	  2008 Jun	&	19	& 39.4	&38.9,40.2 & 8.6\\
G\,12.904--0.031 	&	18 13 48.27 	&	--17 45 38.8 	&	  2008 Jun	&	0.4 	&	59.3	& 59.3,59.5	&	0.1 \\
				&				&				&	  2008 Dec	&	0.6 	&	 59.2 &58.9,59.7 	&	0.3	\\
G\,12.909--0.260$^{1,2,5,7,8}$  	&	18 14 39.53 	&	--17 52 00.0 	&	  2008 Jun	&	15	& 39.4 	& 38.5,40.5	&	14\\
G\,13.696--0.156 	&	18 15 51.05 	&	--17 07 29.6 	&	  2008 Jun	&	0.7 	&	99.2 	&99.0,105.2 &0.6 \\
				&				&				&	  2008 Dec	&	0.6 	&99.2 	& 98.7,105.1 &	0.4\\
G\,14.101+0.087$^{9}$ 	&	18 15 45.81 	&	--16 39 09.4 	&	  2008 Dec	&	11	&	11.2 &	5.7,16.6	&11 \\
G\,14.390--0.020 	&	18 16 43.77 	&	--16 27 01.0 	&	  2008 Dec	&	0.8	&	28.1 &	27.8,28.2 &0.2 \\
G\,14.521+0.155 	&	18 16 20.73 	&	--16 15 05.5 	&	 2008 Dec		&	 0.4 	& 2.1	&1.9,2.1 & 0.1 \\
				&				&				&	 2010 Mar		&	$<$0.79	&	&	--140,200\\
G\,14.604+0.017$^{1}$  	&	18 17 01.14 	&	--16 14 38.0 	&	 2008 Jun		&	 0.6 	&24.7 	& 24.6,24.8	&	0.1\\
G\,15.034--0.677$^{1,2,5,7,8}$  	&	18 20 24.78 	&	--16 11 34.6 	&	  2008 Jun	&	 21 	&23.4 	& 21.2,23.7 & 12 \\
G\,15.094+0.192 	&	18 17 20.82 	&	--15 43 46.5 	&	  2008 Dec	&	1.2 		& 23.4 	& 22.9,23.9 & 0.7\\
G\,16.302--0.196 	&	18 21 07.83 	&	--14 50 54.6 	&	  2008 Dec	&	6.3 		&51.4 	& 51.0,52.5 	&5.2 \\
G\,16.585--0.051$^{1,2}$  	&	18 21 09.13 	&	--14 31 48.5 	&	  2008 Jun	&	2.2		&65.9 	&	59.2,68.7	&2.1 \\
G\,16.976--0.005 	&	18 21 44.68 	&	--14 09 48.5 	&	  2008 Dec	&	0.4 		& 6.6 	&6.5,6.7 &0.1\\
				&				&				&	  2010 Mar	&	0.4		& 6.5 	& 6.5,6.7 & 0.1\\
G\,17.862+0.074 	&	18 23 10.10 	&	--13 20 40.8 	&	  2008 Dec	&	0.5		& 108.2 	&107.9,119.7 &0.5  \\
G\,18.073+0.077 	&	18 23 33.98 	&	--13 09 25.0 	&	  2008 Dec	&	0.5 		& 57.2 	& 46.6,57.3 	&0.5 \\
G\,18.262--0.244 	&	18 25 05.70 	&	--13 08 23.2 	&	  2008 Dec	&	5.2 		&79.1 	&74.5,79.4 &3.0 \\
G\,18.341+1.768 	&	18 17 58.13 	&	--12 07 24.8 	&	  2008 Dec	&	3.5 		&28.0 	&	27.4,29.0 & 1.5 \\
\end{tabular}\label{tab:12MMB}
\end{table*}

\begin{table*}\addtocounter{table}{-1}
  \caption{-- {\emph {continued}}}
  \begin{tabular}{lllllccclclllclllcll} \hline
\multicolumn{1}{c}{\bf Methanol maser} &\multicolumn{1}{c}{\bf RA} & \multicolumn{1}{c}{\bf Dec}  &{\bf Epoch}&  {\bf S$_{12.2}$} & {\bf Vp$_{12.2}$} & {\bf Vr$_{12.2}$} & \multicolumn{1}{c}{\bf I$_{12.2}$} \\
    \multicolumn{1}{c}{\bf ($l,b$)}&  \multicolumn{1}{c}{\bf (J2000)} & \multicolumn{1}{c}{\bf (J2000)}   & & {\bf (Jy)} & {\bf (\kmsns)} & {\bf  (\kmsns)} & \multicolumn{1}{c}{\bf (Jy \kmsns)}\\
    \multicolumn{1}{c}{\bf (degrees)}  & \multicolumn{1}{c}{\bf (h m s)}&\multicolumn{1}{c}{\bf ($^{o}$ $'$ $''$)}\\  \hline \hline
G\,18.667+0.025 	&	18 24 53.78 	&	--12 39 20.4 	&	  2008 Dec	&	0.7 		&76.5	&76.3,76.6 & 0.2 \\
				&				&				&	  2010 Mar	&	$<$0.5	&	&--140,200	\\
G\,18.874+0.053 	&	18 25 11.34 	&	--12 27 36.8 	&	  2010 Mar	&	0.6 		&38.5	& 38.4,38.7 & 0.2 \\
G\,19.009--0.029 	&	18 25 44.78 	&	--12 22 46.1 	&	  2008 Dec	&	0.5 		& 55.8 	& 55.8,58.8 & 0.1 \\
G\,19.249+0.267 	&	18 25 08.02 	&	--12 01 42.2 	&	  2008 Dec	&	 0.4 		&	13.3 & 13.3,13.4 &0.1\\
G\,19.365--0.030 	&	18 26 25.79 	&	--12 03 52.0 	&	 2008 Dec		&	 1.0 		& 29.8 	& 24.8,30.0 & 0.5 \\
G\,19.472+0.170n 	&	18 25 54.72 	&	--11 52 33.0 	&	  2008 Dec	&    0.6 		&	21.5 &20.5,22.5 & 0.4 \\
				&				&				&	2010 Mar		&	0.4 		& 22.2 & 20.6,22.3 & 0.2 \\
G\,19.486+0.151$^{2}$ 	&	18 26 00.39 	&	--11 52 22.6 	&	  2010 Mar	&   2.5 	&	20.7 & 20.4,26.6 & 2.1 \\
G\,19.612--0.134$^{1,2}$  	&	18 27 16.52 	&	--11 53 38.2 	&	  2008 Jun	&	1.4 	& 50.5	&50.2,55.1 & 0.8 \\
G\,19.614+0.011 	&	18 26 45.24 	&	--11 49 31.4 	&	  2008 Dec	&	1.0 	&	34.7 &34.6,34.8 & 0.2 \\
G\,19.667+0.114 	&	18 26 28.97 	&	--11 43 48.9 	&	 2010 Mar		& 	0.6 	&	14.3 & 14.2,16.8 & 0.2  \\
\hline

\end{tabular}
\end{table*}

Spectra of each of the detected 12.2-GHz methanol masers are presented in Fig.~\ref{fig:12MMB}. In the top left hand corner of each spectrum, the epoch of the observations is noted. In the case where the spectrum is the average of multiple epochs, each of those epochs is listed. Spectra are presented in order of increasing Galactic longitude with the exception of nearby sources which required vertical alignment to show features that were also detected at nearby positions. A velocity range of 30~\kms is shown, and is centred on the velocity of the 6.7-GHz methanol maser peak. Two epochs are presented for G\,19.472+0.170n to highlight the reliability of weak features detected at a single epoch when the 6.7-GHz source velocity is well matched by weak 12.2-GHz emission.

\begin{figure*}
	\psfig{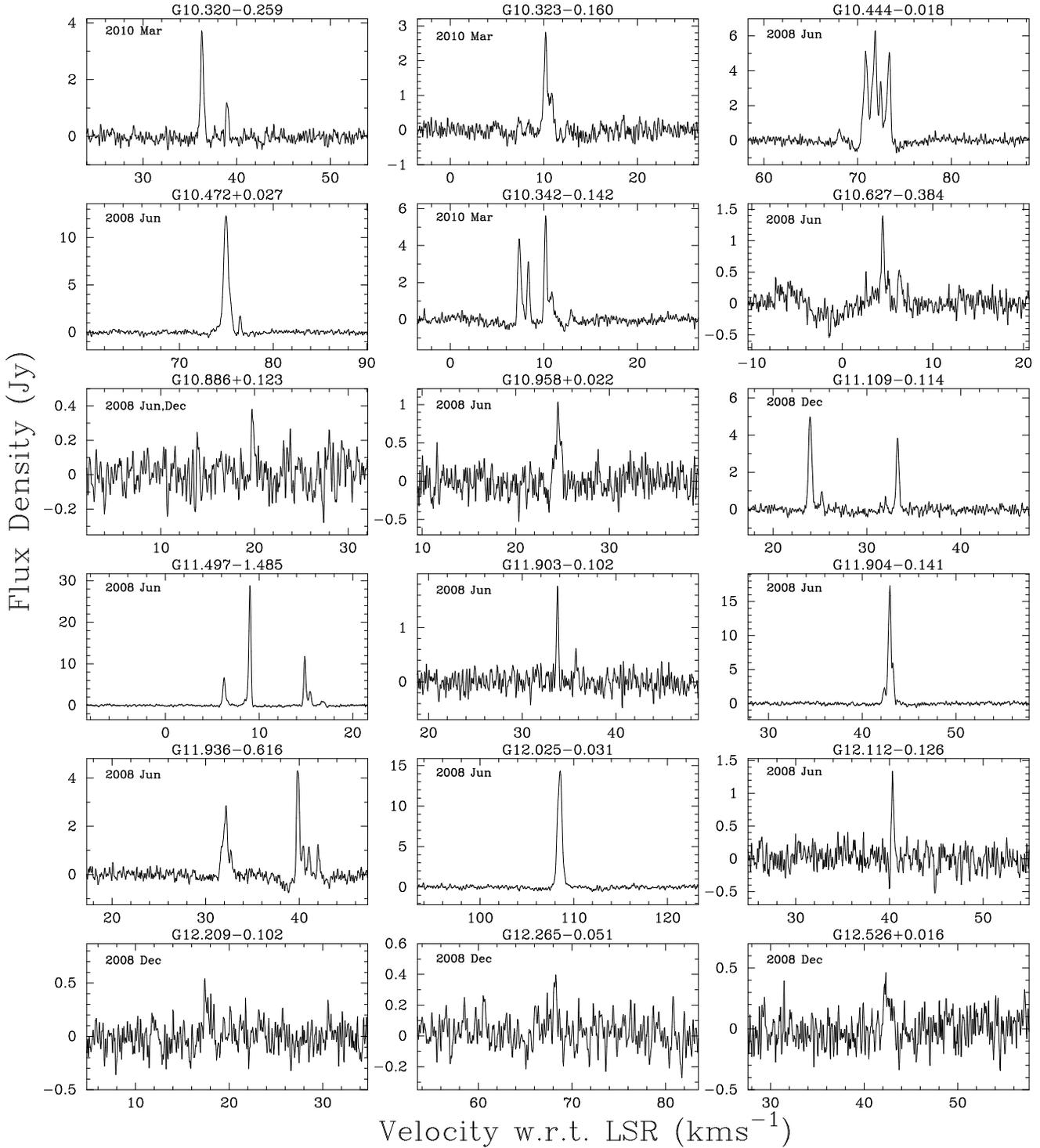}
\caption{Spectra of the 12.2-GHz methanol masers detected towards 6.7-GHz MMB sources.}
\label{fig:12MMB}
\end{figure*}

\begin{figure*}\addtocounter{figure}{-1}
	\psfig{figure=12mmb_paper2.ps}
	\caption{--{\emph {continuued}}}
\end{figure*}

\begin{figure*}\addtocounter{figure}{-1}
	\psfig{figure=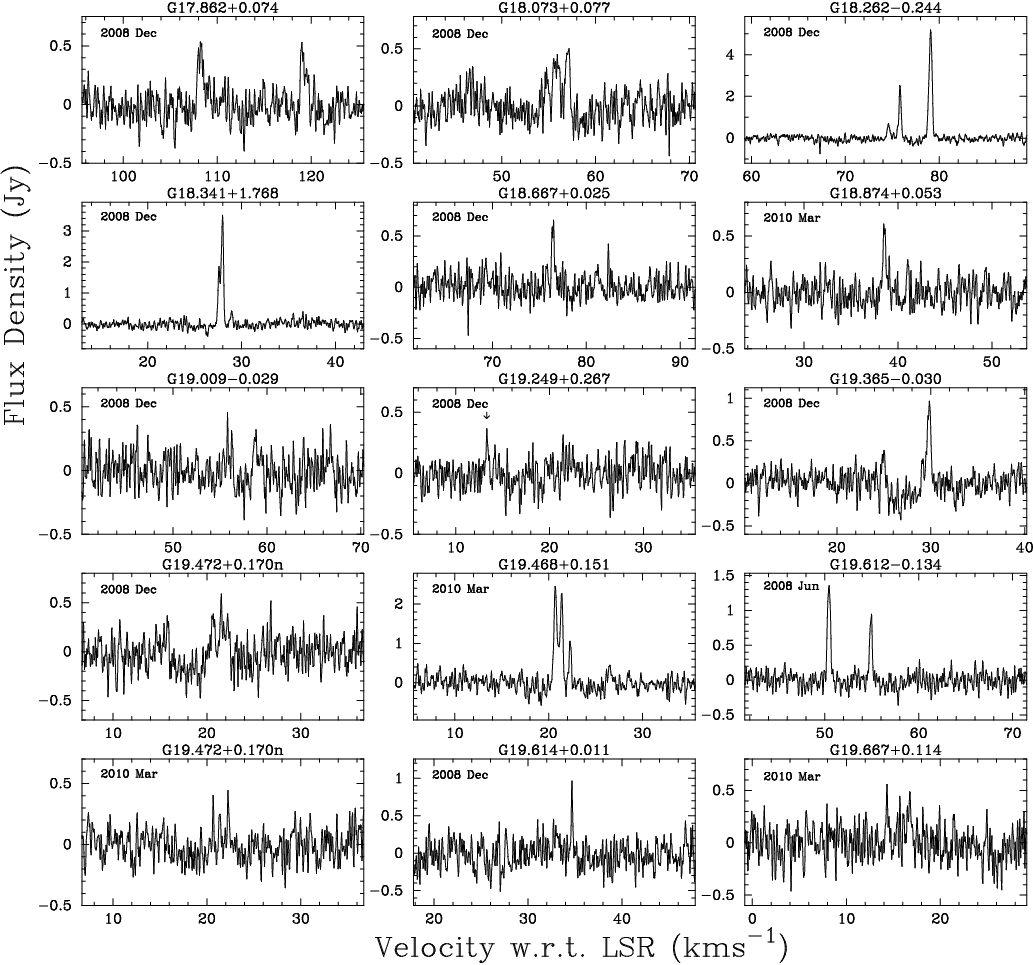}
\caption{--{\emph {continuued}}}
\end{figure*}

\section{Individual source comments}\label{sect:ind}

%

%
%
%

In this section we highlight sources that are not able to be completely described in Tables~\ref{tab:6MMB} and~\ref{tab:12MMB} and Fig.~\ref{fig:12MMB}. In particular, sources which are located within small angular separations, sources that do not formally meet a 5-$\sigma$ detection limit, or have associations with notable objects, are discussed. Comments relating to smoothing are in addition to the initial Hanning smoothing applied to the data as part of the initial processing. 

{\em G\,10.323--0.160 and G\,10.342--0.142.} This close pair of 12.2-GHz methanol maser detections have been vertically aligned in Fig.~\ref{fig:12MMB} to highlight the contributions to each spectrum from the other source. Given that the additional spectral features associated with G\,10.342--0.142 at about 9~\kms are weak in the G\,10.323--0.160 spectrum, it is clear that the contribution to the emission peak at 10.2~\kms is low.

{\em G\,10.472+0.027.} This well studied 12.2-GHz methanol maser is associated with the highest velocity water maser emission detected towards a high-mass star formation region, with red-shifted emission 250~\kms from the systemic velocity \citep{Tit13}. The middle of the 6.7-GHz velocity range \citep[67.5 \kmsns;][]{GreenMMB10} well matches that of the ammonia emission \citep[66.8 to 68.8 \kmsns;][]{Churchwell1990,Purcell12} whereas the 12.2-GHz emission is restricted to slightly red-shifted velocities (73.6 to 76.7 \kmsns).

{\em G\, 10.627--0.384.} Although of a reduced magnitude, we note the presence of real absorption evident in this spectrum (despite our observing strategy), similar to that detected previously in the observations of \citet{Caswell95b}.

{\em G\,10.886+0.123.} The 12.2-GHz spectrum presented in Fig.~\ref{fig:12MMB} is the average of the 2008 Jun and 2008 Dec epochs, highlighting the weak single feature at 19.7~\kms that was visible in the 2008 Dec spectrum. Since the signal-to-noise in the averaged spectrum is higher than the 2008 Dec spectrum, weak emission must have been present during both observation epochs. An additional smoothing over two channels has been applied to the presented spectrum.

{\em G\,11.936--0.616.} Despite an overall loss in flux density, this source remains remarkably similar to the spectrum given in \citet{Caswell95b}, including the small absorption feature to the negative side of the peak emission.

{\em G\,12.265--0.051.} Very weak emission, not detected during 2008 Jun observations, was identified in the 2008 Dec observations. The spectrum shown in Fig.~\ref{fig:12MMB} has been smoothed over an additional two channels and shows a 0.4~Jy peak at the velocity of the 6.7-GHz methanol maser peak, and an exceptionally weak secondary feature at $\sim$60.5~\kmsns, corresponding in velocity to a 6.7-GHz spectral feature of about 0.7~Jy.

{\em G\,14.457-0.143.} We note that there is a typographical error in the 6.7-GHz maser position listed in the MMB catalogue \citep{GreenMMB10}.  Its Galactic coordinates are given correctly, but the declination given is --16\degr 27\arcmin 57\farcs5, whereas, the correct declination is --16\degr 26\arcmin 57\farcs5.  The 12.2-GHz observations were conducted towards the position reported in the in the MMB catalogue, placing it at approximately the half-power point of the 12.2-GHz Parkes beam. This is reflected in the uncharacteristically high 5-$\sigma$ detection limit listed in Table~\ref{tab:6MMB}.

{\em G\,18.661+0.034 and G\,18.667+0.025.} We detect a single 12.2-GHz methanol maser feature towards this close pair of sources. The velocity of the 12.2-GHz emission at 76.5~\kms is consistent with the peak 6.7-GHz methanol maser feature associated with G\,18.667+0.025 which has a velocity of 76.6~\kmsns. \citet{GreenMMB10} note that there is some contribution at this velocity from G\,18.661+0.034, but that the majority of the emission comes from G\,18.667+0.025. Although we can not rule out that some (or perhaps all) of the 12.2-GHz emission may be associated with G\,18.661+0.034, it seems unlikely given that there is no 12.2-GHz emission associated with the stronger 6.7-GHz spectral features associated with this source and that the contribution from G\,18.661+0.034 to the spectral feature at 76.6~\kms is small. We therefore attribute the 12.2-GHz emission to G\,18.667+0.025.


{\em G\,19.009--0.029.} A weak 12.2-GHz detection observed at a single epoch (2008 Dec) with a ten minute integration resulting in a 5-$\sigma$ noise level of 0.56~Jy. The peak of the detected emission at 0.4 Jy makes it a 3.6-$\sigma$ detection. The velocity of the narrow 12.2-GHz peak (at 55.8~\kmsns) falls within the strongest emission associated with the 6.7-GHz source, but a weaker, much broader 12.2-GHz emission at $\sim$59~\kms adds credibility to this source as it, too, matches the velocity of emission associated with the 6.7-GHz source.

{\em G\,19.249+0.267.} We detected a single-feature, 0.4 Jy (3.6-$\sigma$) source at a velocity of 13.3 \kmsns, which corresponds to the velocity of the third strongest 6.7-GHz feature ($\sim$1.8~Jy), during the only observation made at this position in 2008 Dec.

{\em G\,19.667+0.114.} Although not a 5-$\sigma$ detection, this weak 12.2-GHz methanol maser shows multiple features at the velocities of corresponding 6.7-GHz methanol maser emission.

\section{Discussion}

This is the third instalment of a series of 12.2-GHz methanol maser catalogues, comprising the largest, statistically complete sample of these methanol masers ever compiled. Although targeted towards 6.7-GHz methanol masers, such a search is likely to capture the entire population of 12.2-GHz methanol masers since there have been no serendipitous detections of 12.2-GHz methanol masers without 6.7-GHz counterparts to date, and, furthermore, 12.2-GHz methanol masers only rarely have larger flux densities than their 6.7-GHz counterparts \citep[e.g.][]{Gay,Caswell95b,Blas04,BreenMMB12a,BreenMMB12b}. Therefore the unbiased manner in which the MMB survey was conducted, makes this series of catalogues an excellent representation of the entire 12.2-GHz methanol maser population.

In addition to the two previous 12.2-GHz catalogues \citep{BreenMMB12a,BreenMMB12b}, the properties of the 12.2-GHz masers detected towards the complete sample of MMB masers that lie south of declination --20$^{\circ}$ (580 sources) have been derived \citep{Breen12stats}. These have shown that 12.2-GHz methanol masers are typically found towards the more luminous 6.7-GHz methanol masers, which are interpreted as being the more evolved objects, and therefore it is expected that the 12.2-GHz sources trace the second half of the 6.7-GHz methanol maser lifetime. The 12.2-GHz methanol maser luminosity is also expected to increase with evolution, although at a slower rate than its 6.7-GHz counterpart. This trend was shown to hold among multiple spectral features associated with single sources \citep{BreenMMB12a}, with the scatter in the ratio of 6.7-GHz to 12.2-GHz emission of single features much less within a single source than across the entire sample. 

Within the catalogue papers \citep{BreenMMB12a,BreenMMB12b} we have also investigated such things as the completeness of our 12.2-GHz search, the Galactic distribution of 6.7-GHz sources with and without 12.2-GHz emission, temporal variability of 12.2-GHz emission and the incidence of rarer class II methanol maser transitions. In this third instalment, we focus our investigation on the association with water and OH masers and GLIMPSE sources and their impact on our evolutionary scenario, complementing those investigations previously presented.

The basic detection statistics are, not surprisingly, similar in the 10$^{\circ}$ to 20$^{\circ}$ longitude range to those presented previously. Within this range, we detect 12.2-GHz methanol masers towards 47 per cent of 6.7-GHz sources, compared to 40 percent for the 186$^{\circ}$ to 330$^{\circ}$ range \citep{BreenMMB12b} and 46 per cent for the 330$^{\circ}$ (through 0$^{\circ}$) to 10$^{\circ}$ longitude range \citep{BreenMMB12a}. Our overall detection rate of 45 per cent is lower than those found in earlier searches \citep[for which the 6.7-GHz targets were biased towards those associated with OH masers and {\em IRAS} sources; see discussion in][]{Breen12stats}. It is also noted in \citet{Breen12stats} that the 12.2-GHz detection rate is not constant over the range of Galactic longitudes, perhaps due to a metallicity gradient across the Galactic plane, or possibly dominated by localised bursts of star formation that have resulted in many nearby sources being at a similar evolutionary stage.

Within the current Galactic longitude range, we detect 12.2-GHz methanol masers that range in peak flux density from 0.4 to 29~Jy and the majority of sources (27) have peak flux densities less than 2~Jy. The velocity extent of the detected sources range from 0.1 to 11.8~\kmsns. Their spread of peak flux densities and velocity ranges are confined to smaller values than those sources in the previous catalogue sections (which were dominated by strong sources of which there are few in the much smaller longitude range considered here).

\begin{figure}\vspace{-1.5cm}
\begin{center}
	\epsfig{figure=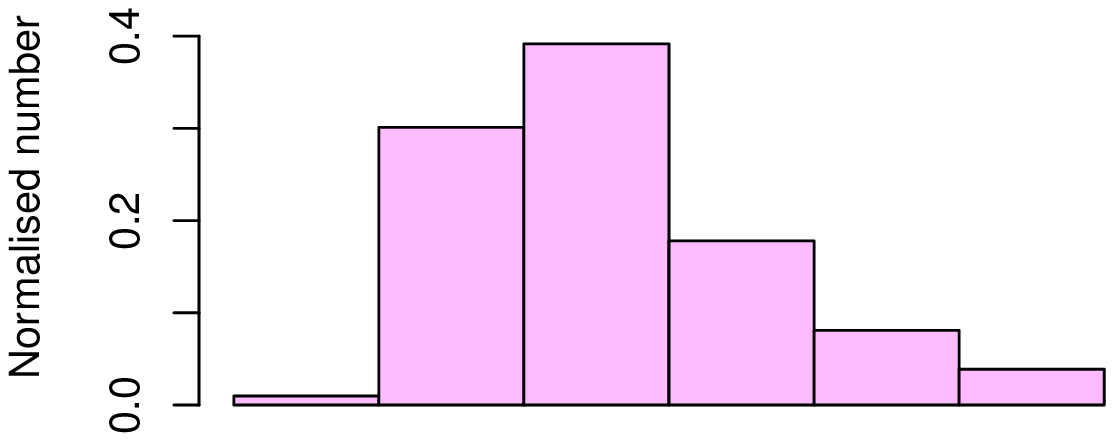,height=9cm,angle=270}\vspace{-3cm}
	\epsfig{figure=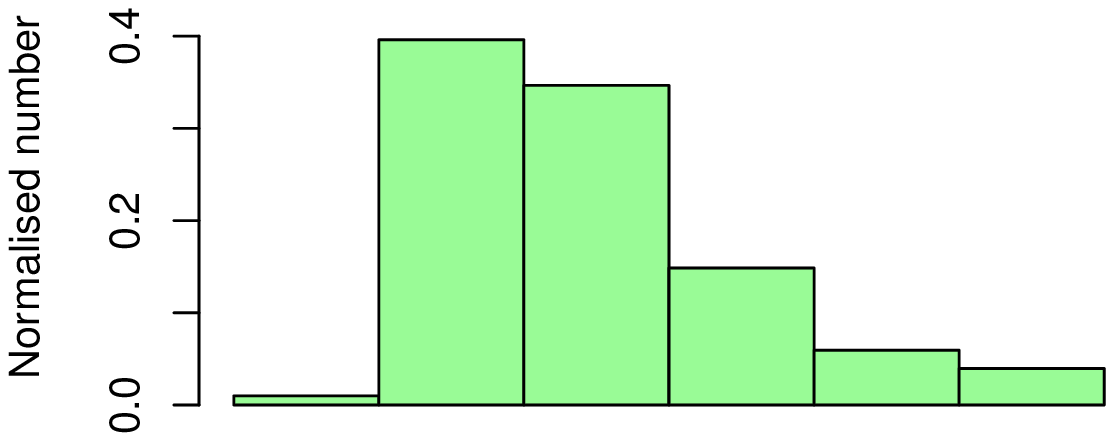,height=9cm,angle=270}\vspace{-3cm}
	\epsfig{figure=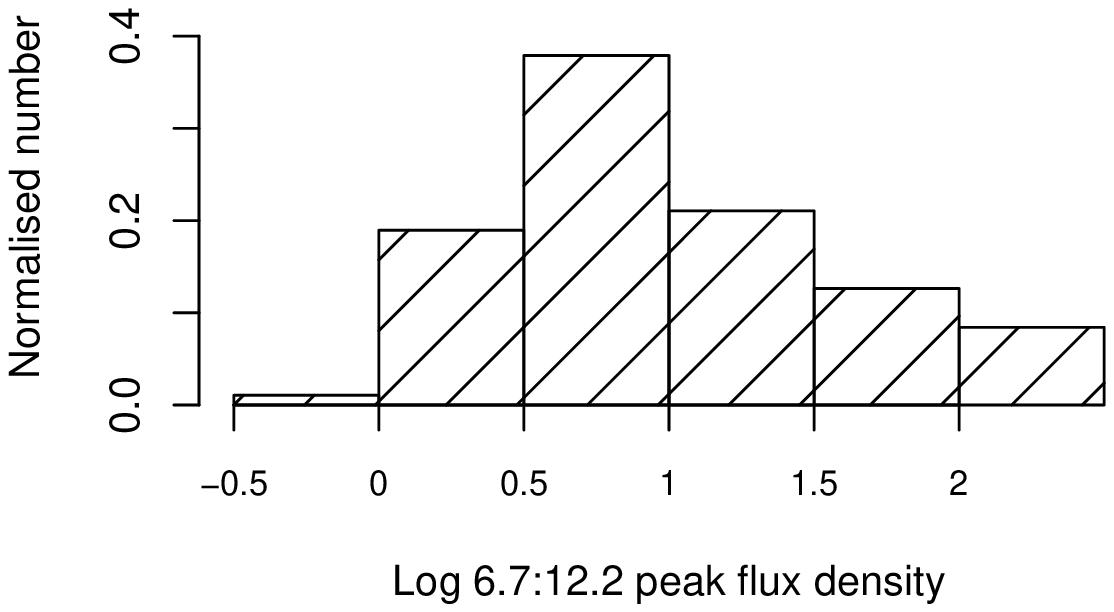,height=9cm,angle=270}
\caption{Histogram of the log 6.7- to 12.2-GHz peak flux density ratio for all sources in the 186$^{\circ}$ to 20$^{\circ}$ longitude range (mauve; panel 1), sources with no OH masers [green; panel 2 (incorporating all sources covered by the \citet{C98} complete search for OH masers)], and sources with associated OH masers [hashed; panel 3 \citep[][]{C98}]. Each panel has been normalised relative to the total number of members in each category (309 - all sources; 101 - sources without OH; 95 - sources with OH).}
\label{fig:612hist}
\end{center}
\end{figure}

Fig.~\ref{fig:612hist} shows histograms of the 6.7- to 12.2-GHz methanol maser peak flux density for all sources over the longitude range 186$^{\circ}$ to 20$^{\circ}$, as well as the subsamples of sources with and without associated OH masers \citep{C98}. The values shown in each histogram have been normalised by the number of sources in each of the three categories to allow the distributions to be seen without the difficulties introduced by different sample sizes. The histograms showing sources with and without OH masers nicely represent a result found previously - that the ratio of 6.7- to 12.2-GHz emission changes with source evolution, interpreting those OH-associated sources to be more evolved \citep{FC89,C97}. In particular, Fig.~\ref{fig:612hist} shows that sources associated with OH maser emission are skewed towards larger values of 6.7- to 12.2-GHz peak flux density ratio than those without OH masers: reflected in the results of a Kolmogorov-Smirnov test which gave a p-value of 0.001, allowing the null hypothesis (that they come from the same distribution) to be rejected.



\subsection{Comparison with water maser emission}\label{sect:water}

%

\begin{figure}\vspace{-1.5cm}
\begin{center}
	\epsfig{figure=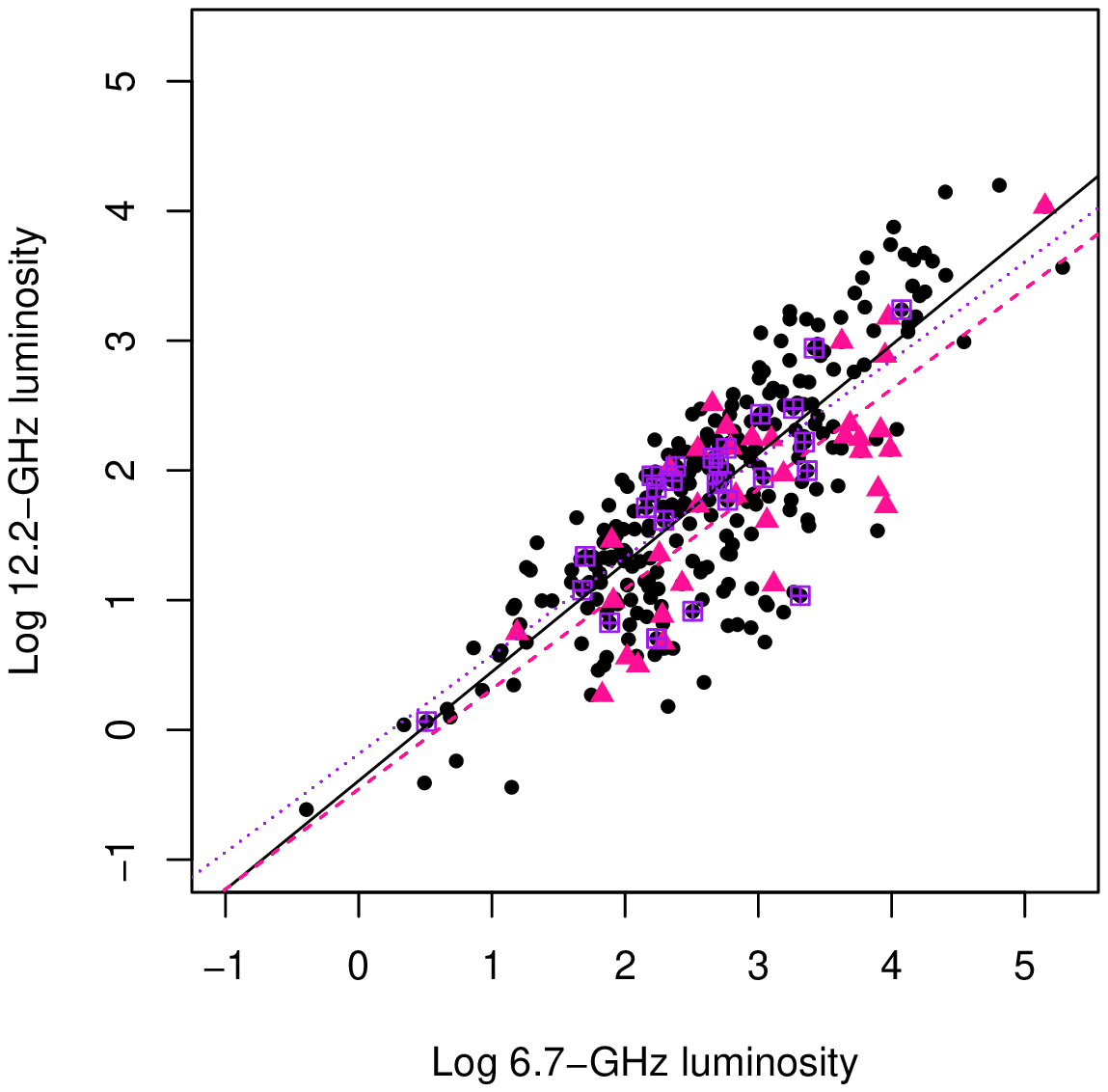,height=9cm,angle=270}
\caption{Peak luminosity of the 12.2-GHz maser emission vs the peak luminosity of the 6.7-GHz maser emission (units of Jy kpc$^2$) for 281 of the 314 (some have been excluded due to poor distance determinations or due to confusion with nearby sources)12.2-GHz methanol masers detected in the longitude range 186$^{\circ}$ (through the Galactic centre) to 20$^{\circ}$  \citep[][plus the current paper]{BreenMMB12a,BreenMMB12b} (black dots). The pink triangles represent sources with an associated water maser \citep{Titmarsh13} in the 6$^{\circ}$ to 20$^{\circ}$ longitude range. The black dots that are encased in a purple square represent the population of 6.7 and 12.2-GHz methanol masers covered by the \citep{Titmarsh13} water maser search where no detection was made. }
\label{fig:lum}
\end{center}
\end{figure}

The recent water maser follow-up to the MMB targets covering the Galactic longitude range 6$^{\circ}$ to 20$^{\circ}$ \citep{Titmarsh13} has allowed us to compare the occurrence of 12.2-GHz methanol masers and water masers towards a complete sample of 6.7-GHz methanol masers for the first time. Combining the current catalogue with the methanol masers in the 6$^{\circ}$ to 10$^{\circ}$ longitude range also covered in the \citet{Titmarsh13} water maser follow-up observations, we find that of the 119 6.7-GHz methanol masers in this longitude region, 61 have associated 12.2-GHz methanol masers. Of the 55 water masers detected towards the 119 6.7-GHz methanol masers in this range \citep{Titmarsh13}, 34 are associated with sources showing 12.2-GHz emission, and 21 are associated with sources devoid of detectable 12.2-GHz methanol maser emission. Water masers are therefore detected towards a larger fraction of 6.7-GHz methanol masers with 12.2-GHz emission (55.7 per cent) than those without (36.2 per cent). 

Water masers are often noted for their sometimes extreme temporal variability \cite[e.g.][]{Brand2003,Felli07}. While the level to which this affects the detection of sources with 6.7-GHz methanol masers is not well characterised, we can estimate the number of water masers that would have been missed in the \citet{Titmarsh13} observations using a two epoch study of water masers towards a sample of methanol masers \citep{Breen10b}. In their study, \citet{Breen10b} targeted  a number of sources with methanol masers, finding water maser emission towards 132 sources on at least one of the two observation epochs (separated by 10 months). Of the 132 sources, 15 fell below the 0.2~Jy detection limit on one of the two epochs, implying that $\sim$5.5 per cent of water masers at the sites of 6.7-GHz methanol masers are missed in single-epoch searches. This implies that $\sim$3 sources of water maser emission were missed in the \citeauthor{Titmarsh13} sample due to variability. Given that this is a small number of sources we can conclude that water maser variability is unlikely to have a significant effect on our detection statistics or subsequent analysis.

Previous analysis of 6.7-GHz methanol maser data from the MMB survey and 12.2-GHz follow-up observations \citep{Breen10a,Breen12stats} found that the luminosity of the 6.7-GHz and 12.2-GHz methanol masers increase as the sources evolve (although at a slower rate in the case of the 12.2-GHz emission) and that the 12.2-GHz sources were associated with the later stages of the 6.7-GHz methanol maser lifetime. We have therefore investigated the 6.7-GHz and 12.2-GHz luminosity properties of the sources with and without associated water maser presence. Fig.~\ref{fig:lum} shows the peak 12.2-GHz methanol maser luminosity versus the associated 6.7-GHz peak luminosity, depicted by black dots for the whole sample. On the plot, methanol sources in the 6$^{\circ}$ to 20$^{\circ}$ longitude range with associated water masers are represented by pink triangles, and sources without associated water maser emission are highlighted with purple crossed-squares. Also overlaid are the line-of-best-fit for each of the three groups, colour-coded to match the symbols. These lines are equivalent to within the errors in both slope and intercept, confirming what is evident from basic visual inspection - there is no appreciable difference in the 6.7- and 12.2-GHz methanol maser peak luminosity properties between the groups. 

The results of similar comparisons with the presence of rarer class II methanol masers have shown very different results. \citet{Ellingsen11} found that the presence of 107-GHz methanol masers was limited to sources with high 6.7-GHz luminosities, while 37-GHz methanol masers were associated with sources showing both high 6.7- and 12.2-GHz methanol maser luminosities (see their fig. 4), a relationship further investigated and confirmed by \citet{Ellingsen13}. Combining this with the expected evolution of the 6.7-GHz and 12.2-GHz methanol maser luminosity led \citet{Ellingsen11,Ellingsen13} to suggest that the 37.7-GHz methanol masers likely trace a short-lived phase signifying the end of the phase where class II methanol masers are seen, slightly preceded by the onset of the 107-GHz methanol maser emission. In contrast, the sources associated with water masers show no systematic trends in 6.7-GHz and 12.2-GHz methanol maser luminosities (but note that by selection they are generally associated with the more luminous 6.7-GHz methanol masers), even though we find that water masers are more likely to be detected towards sources with 12.2-GHz emission. Perhaps this is reflecting the different pumping mechanisms of the respective masers: water masers are collisionally pumped and are often observed towards sources with outflows, a property which may be of greater prevalence towards older sources as traced by 12.2-GHz methanol masers.  The general success of the maser-based evolutionary timeline in producing a self-consistent explanation of class II methanol and OH masers may be because those transitions are radiatively pumped by the dust emission which is warmed by the protostar and hence more closely tied to its properties.  In contrast, the collisionally pumped maser transitions such as water (and perhaps also class I methanol) depend on interactions between the protostar and the environment (e.g. the presence of outflows interacting with molecular gas).  This is consistent with the large number of water masers throughout the Galaxy, as if they simply require an outflow and molecular gas those conditions may exist over longer timescales than do the appropriate radiative conditions close to the protostar.  It also explains why water masers cannot easily be tied to a specific evolutionary phase, since for some protostars in the class II maser phase they may have an outflow interacting with surrounding molecular gas, or they may not.  The probability of interaction may generally evolve over time (as the power of the outflow evolves), but this is not likely to produce a good correlation between maser occurrence and evolutionary phase.

\subsection{Comparison with OH masers}

\citet{C98} presents a large catalogue of 1665- and 1667-MHz OH masers, predominantly detected towards sites of star formation. It incorporates the results of a complete search of a strip of the Galactic plane, along with targeted observations of known sources, conducted with the ATCA and resulting in precise positions of more than 200 1665-MHz OH masers. Using the 6.7-GHz methanol maser data available at the time, \citet{C98} confirmed the close association of OH masers with 6.7-GHz methanol masers, estimating that 80 per cent of OH masers have 6.7-GHz methanol maser counterparts. The complementary association rate of OH maser towards methanol masers is estimated to be much lower, perhaps less than 40 per cent \citep{Breen12stats}, accounting for the greater numbers of detected methanol masers. Statistics on the population of OH masers detected towards methanol masers, and those without associated methanol masers will be more accurately represented by ongoing surveys for the ground-state OH maser lines \citep[MAGMO, SPLASH:][respectively]{Green12b,Dawson}.

\citet{C96} used the typical range of methanol-to-OH maser peak flux density ratios (R) to introduce the concept of ``methanol-favoured" and ``OH-favoured" maser sites, where methanol-favoured correspond to R $>$ 32 and OH-favoured correspond to sites with R $<$ 8. In a sample of 57 methanol masers, \citet{C96} noted that \UCHII regions were only detected towards three maser sites, all of which were OH-favoured, a criterion met by only 6 sources in the sample. In subsequent analysis of a largely independent sample, \citet{C97} found from a sample of 62 sources that 15/27 OH-favoured maser sites were associated with \UCHII regions, compared to a solitary association between an \UCHII region and a methanol-favoured maser site. \citet{C97} therefore confirmed that stronger \UCHII regions were detected preferentially towards OH-favoured sites, although cautioned this was a loose correlation with a large amount of scatter. \citet{C98} suggested an evolutionary interpretation of these characteristics following a scenario whereby the methanol maser emission was prominent prior to the onset of OH maser emission, the OH and methanol masers coexist at an intermediate phase and  then the emergence of an \UCHII causes the methanol maser emission to fade prior to the OH masers.

\citet{Krishnan13} compared the presence of rarer class II methanol masers to the R values of each of the sites. They found that 107-GHz methanol masers that also have an associated 19.9-GHz methanol masers tended to have lower R values, indicative of OH-favoured, or older sources, while 37-GHz methanol masers tended to be associated with less evolved sources, more often having higher R values. \citet{Ellingsen13} further investigated this trend, similarly noting a preference for 37-GHz methanol masers to be associated with methanol-favoured sources.

We have compared the R values for the large sample of sources given in \citet{C98} with the occurrence and characteristics of corresponding 12.2-GHz methanol masers detected in the longitude range 186$^{\circ}$ (through the Galactic centre) to 20$^{\circ}$ \citep[][and the current paper]{BreenMMB12a,BreenMMB12b}. As expected, a large proportion of the OH-associated 6.7-GHz methanol masers have associated 12.2-GHz methanol masers: of the 163 6.7-GHz methanol maser sites in the \citet{C98} sample that have also been searched for 12.2-GHz emission, 101 (62 per cent) have associated 12.2-GHz methanol maser emission (c.f. 45 per cent for the entire methanol sample, consistent with previous findings that the 12.2-GHz sources are present towards the later stages of the 6.7-GHz methanol maser lifetime, when the presence of an OH maser becomes increasingly likely \citep{FC89,C97}). 

Table~\ref{tab:fav} shows a breakdown of the different 6.7-GHz methanol to OH maser peak flux density ratio categories: OH-favoured, methanol-favoured, or intermediate (corresponding to sources falling in-between the ratio categories) for both 12.2-GHz methanol maser detections and non-detections. The 12.2-GHz detections are fairly evenly spread across the three categories, whereas the 12.2-GHz non-detections are dominated by sources falling in the OH-favoured category. The fraction of 12.2-GHz detections across the R categories shows a clear trend, preferring the methanol-favoured sources  (94 per cent), then the intermediate sources (76 per cent), and, finally, the OH-favoured sources (41 per cent).



\begin{table}
\caption{Sources listed in \citet{C98} that have been searched for 12.2-GHz emission broken up into the categories of OH-favoured (R $<$ 8), methanol-favoured (R $>$ 32) and intermediate (8 $\leq$ R $\leq$ 32) and showing where both the 12.2-GHz detections and non-detections fall.} 
  \begin{tabular}{lcc} \hline
{\bf Source type}  		& {\bf With 12.2-GHz} & {\bf Without 12.2-GHz}\\ \hline
{\bf OH-favoured}		&	34	&	48	\\
{\bf Meth-favoured}		&	29	&	2	\\ 
{\bf Intermediate}		&	38	&	12	\\ \hline
{\bf Total number}		&	101	&	62 \\ \hline
\end{tabular}\label{tab:fav}
\end{table}



\begin{figure}\vspace{-1.5cm}
\begin{center}
	\epsfig{figure=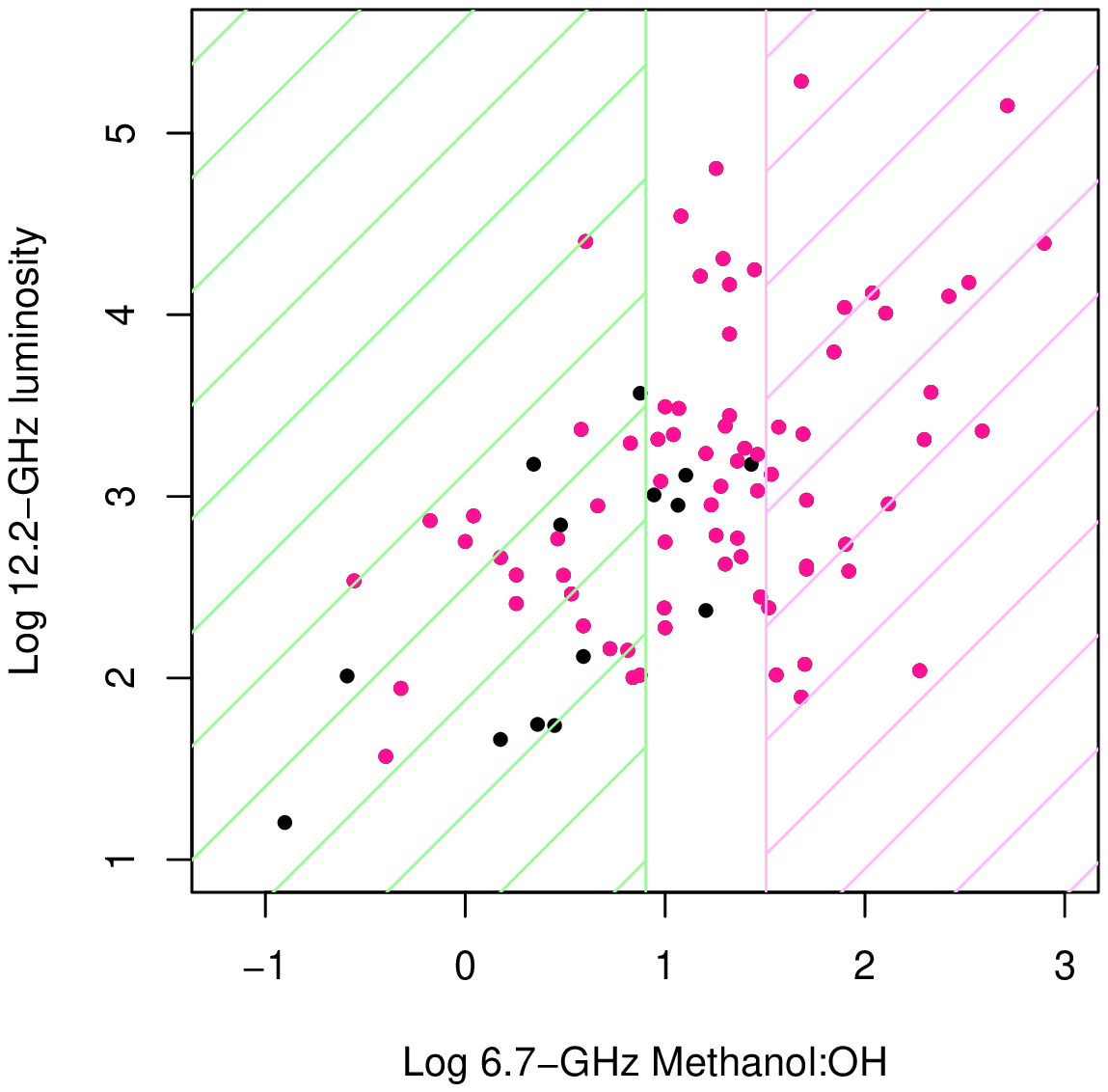,height=9cm,angle=270}
\caption{ Log of the 12.2-GHz peak luminosity (Jy kpc$^2$) versus the log of the 6.7-GHz methanol to OH maser peak flux density as listed in \citet{C98}. Black dots represent those sources where precise positions for the methanol maser emission were not available and as such an association could not be assured \citep{C98}, pink dots represent the majority of the sample where an association had been confirmed. The green shaded area represents the portion of the plot where ``OH-favoured" sources fall (corresponding to R $>$ 8) and the mauve shaded area shows the ``methanol-favoured" section (corresponding to R $>$ 32).}
\label{fig:12lum_R}
\end{center}
\end{figure}

We have compared the 12.2-GHz methanol maser luminosity (calculated using the H{\sc i} self-absorption distances \citep{GM11} or the near kinematic distance using the \citet{Reid09} prescription) with the peak 6.7-GHz to OH ratio for the sample of 163 sources overlapping with the \citet{C98} OH maser sample (shown in Fig.~\ref{fig:12lum_R}). There is a general trend whereby the more luminous 12.2-GHz methanol masers are associated with the sources with larger values of R (tending towards the methanol-favoured sources). Since 12.2-GHz methanol masers are generally detected towards the more luminous 6.7-GHz methanol masers and they are found to increase in luminosity together with evolution, this seems expected. However, initial consideration of the general slope in Fig.~\ref{fig:12lum_R} might seem to suggest an inconsistency between the two evolutionary trends discussed previously: 1) 12.2-GHz methanol masers increase in luminosity as they evolve; and 2) OH-favoured sources (R $<$ 8) are associated with a generally later evolutionary stage of star formation than sources with larger values of R.


 \citet{Breen12stats} compared the luminosities of 6.7- and 12.2-GHz methanol masers with the occurrence of OH maser emission in their fig. 5. The plot shows that for a given 6.7-GHz luminosity the OH-associated sources generally lie at a lower 12.2-GHz luminosity than those sources without OH  maser emission. This may suggest a `turn-over' in the methanol maser luminosity nearing the end of the methanol maser lifetime. It is evident in Fig.~\ref{fig:12lum_R} that the vast majority of sources present on this plot have log 12.2-GHz peak luminosities (Jy kpc$^2$) greater than 2, placing them at the high-end of the 12.2-GHz population [fig. 8 of \citet{Breen12stats} shows the full range of the 12.2-GHz peak luminosities extend to values a factor of $\sim$300 lower than this (to $\sim$-0.5), and the majority have log values less than 2], beyond the luminosity of the majority of sources. An interpretation consistent with this, combined with the comparisons with the R value explored here would be that the 12.2-GHz methanol masers increase in luminosity with evolution (as previously suggested), reach a peak 12.2-GHz luminosity around the time that a weak OH maser becomes detectable, and then as they continue to evolve, the OH maser increases in luminosity, accompanied by a decrease in 12.2-GHz methanol maser luminosity. Undoubtedly, the maximum luminosity attained by the respective maser species is related to the mass of the associated protostar which would be responsible for some of the scatter in plotted luminosities together with the uncertainties in distance determinations.

\begin{figure}\vspace{-1.5cm}
\begin{center}
	\epsfig{figure=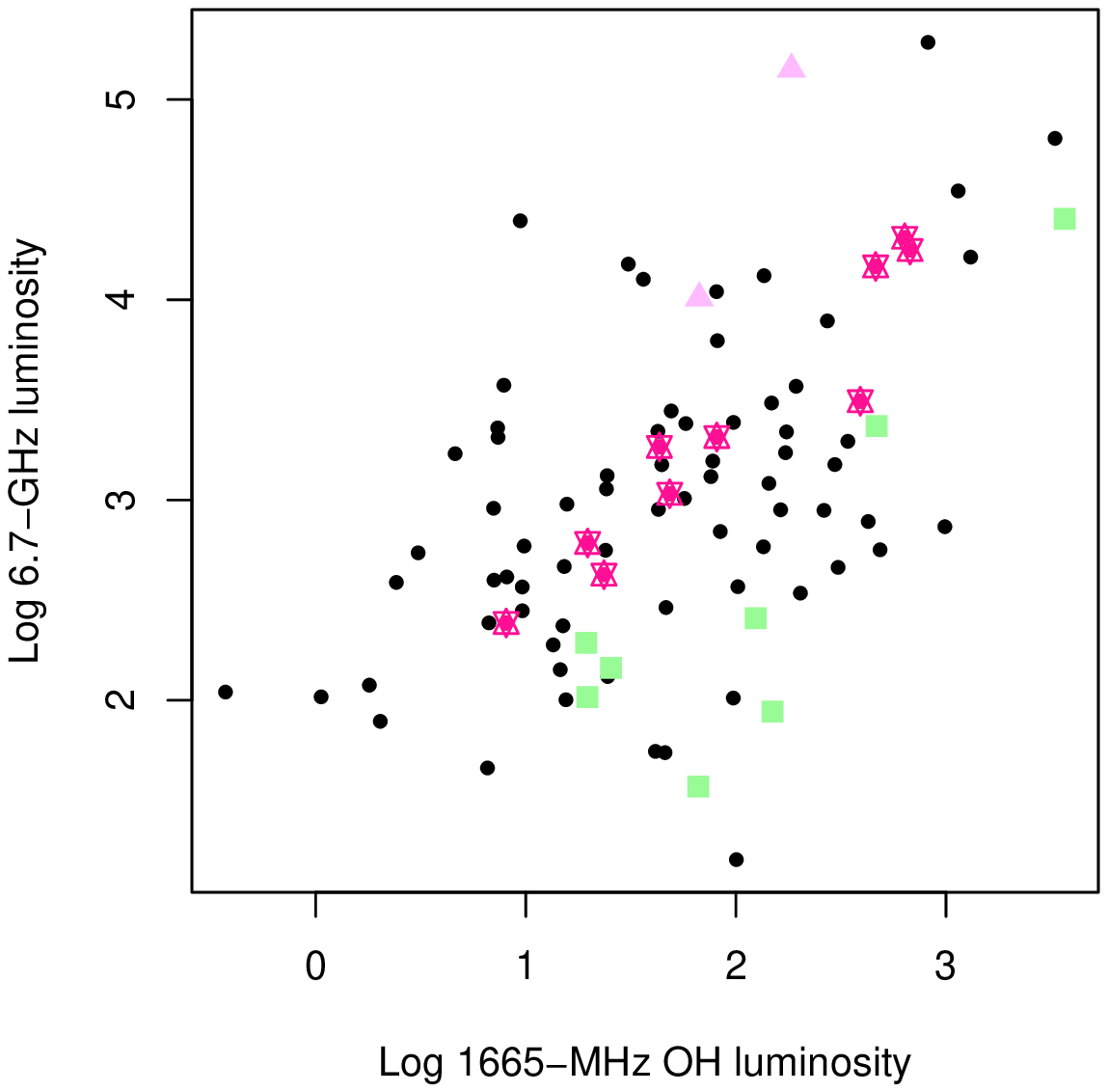,height=9cm,angle=270}
\caption{Log of the peak 6.7-GHz methanol maser luminosities (using peak flux densities taken from the MMB survey data; in units of Jy kpc$^2$) versus log of the 1665-MHz OH maser luminosity \citep[][Jy kpc$^2$]{C98}. The coloured shapes represent those sources where \citet{C98} reports an associated \UCHII region, broken up into OH-favoured (green squares), intermediate (pink stars) and methanol-favoured (mauve triangles). }
\label{fig:lum_cont}
\end{center}
\end{figure}

Fig.~\ref{fig:lum_cont} shows the 6.7-GHz methanol maser luminosity plotted against the 1665-MHz OH maser luminosity and those sources associated with \UCHII regions \citep{C98} have been marked. As expected, few methanol-favoured sites (2) are associated with radio continuum, whereas 10 OH-favoured sites have readily detectable \UCHII region (note that only 8 are plotted on Fig.~\ref{fig:lum_cont} due to the absence of distances for 2 sources).

%

\subsection{Comparison with GLIMPSE emission}

Comparing the occurrence of masers with different GLIMPSE point sources properties has been the focus of several analyses \citep[e.g.][]{Ellingsen06,Breen12stats,Breen10b} using combinations of 6.7- and 12.2-GHz methanol masers, as well as OH and water masers. Each of these investigations has revealed that the maser emission is associated with `extreme' GLIMPSE sources, chiefly lying at values separated from the bulk of a standard set of comparison sources in any combination of colour and magnitude plots. All of these early analyses have been limited to sources that are well characterised as point sources in several, or all, of the GLIMPSE IRAC (Infrared Array Camera) bands. 

\citet{Gallaway13} studied the mid-infrared environments of 776 of the MMB 6.7-GHz methanol masers using the full GLIMPSE survey data; they extended the sample beyond the well-fitted point sources by employing an adaptive noncircular aperture photometry technique to determine the fluxes of the four IRAC bands towards each of the methanol maser sources. Using their larger, and more encompassing sample, they found remarkably similar results to previous large investigations using GLIMPSE point sources; that the sources associated with 6.7-GHz methanol masers are significantly redder than the vast majority of general GLIMPSE population. 

\begin{figure}\vspace{-1.5cm}
\begin{center}
	\epsfig{figure=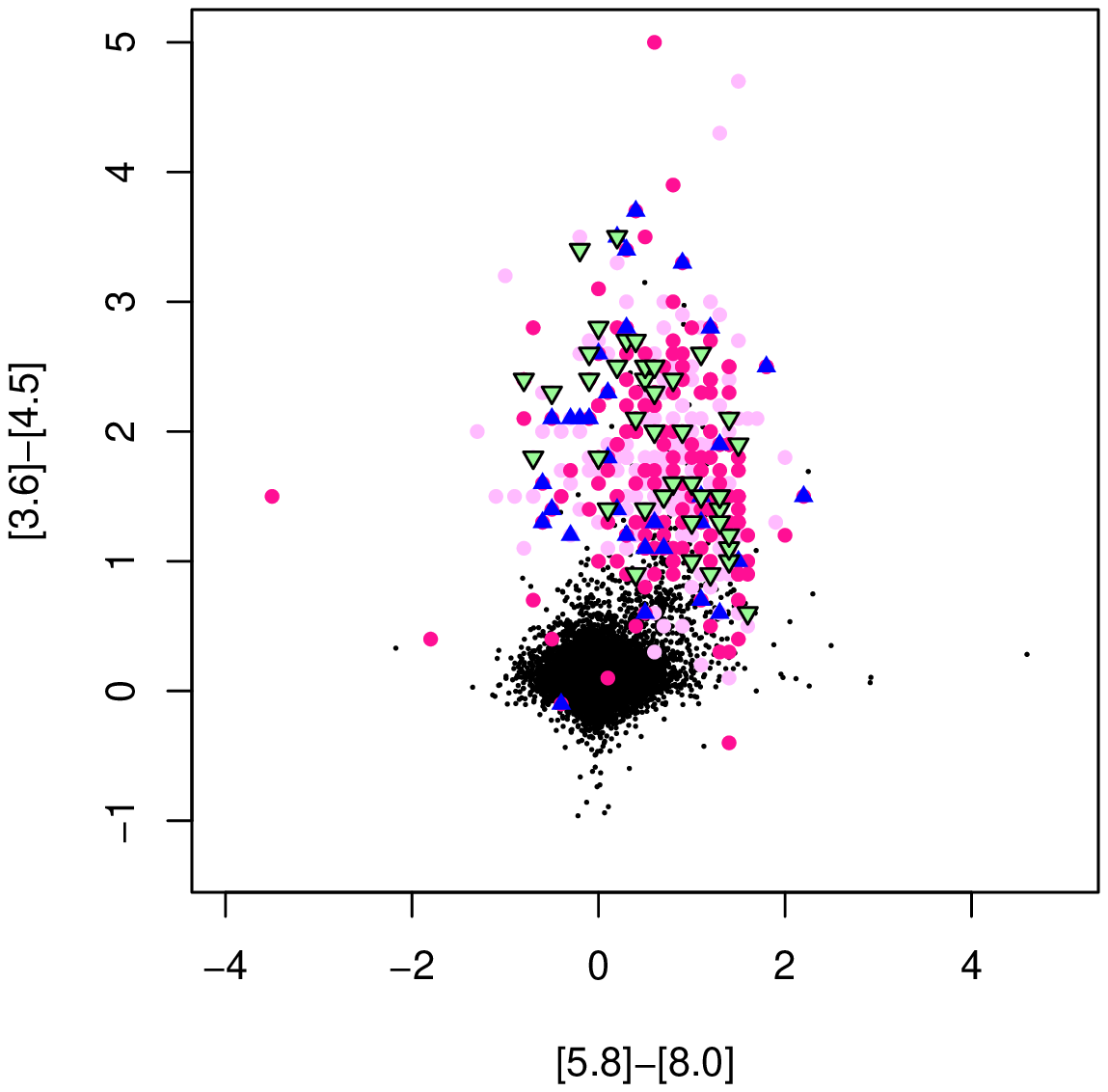,height=9cm,angle=270}
\caption{Colour-colour plot of GLIMPSE data. The black dots represent all of the GLIMPSE point sources within 30 arcmin of $l$ = 326$\fd$5, $b$ = 0$\fd$0. Sources with 12.2-GHz emission (dark pink dots), no 12.2-GHz counterparts (mauve dots), water masers (blue triangles), no water masers (inverted green triangles) have been plotted. Note that only a subset of sources plotted have been searched for water maser emission.  }
\label{fig:col}
\end{center}
\end{figure}

\begin{figure}\vspace{-1.5cm}
\begin{center}
	\epsfig{figure=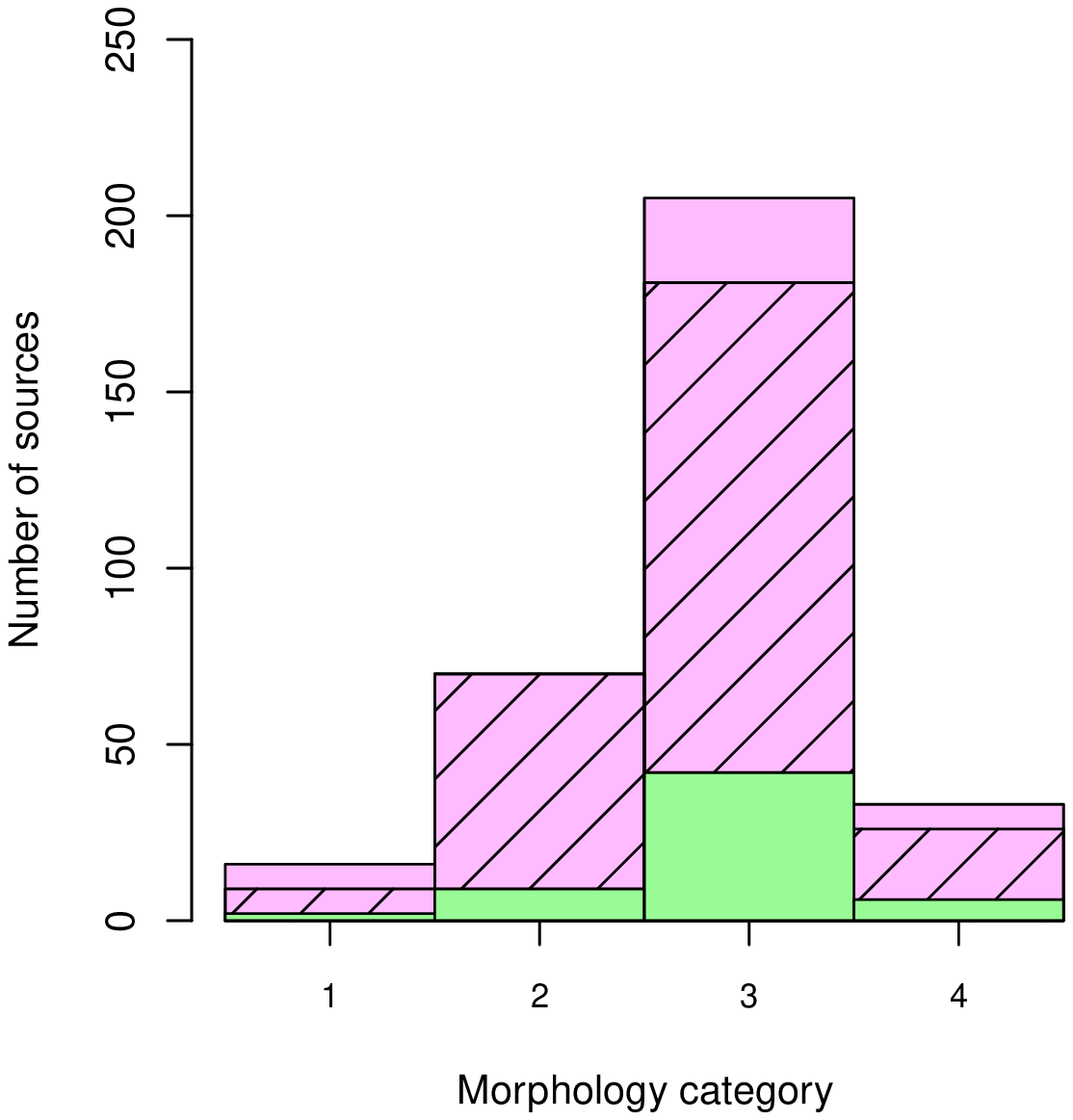,height=9cm,angle=270}
\caption{Distribution of 6.7-GHz methanol masers through the four infrared morphology categories \citep[explained in the text;][]{Gallaway13}. Mauve shows the sources without 12.2-GHz counterparts and overlaid are sources with associated 12.2-GHz emission (black hashing), and green shows the distribution of sources in the 6$^{\circ}$ to 20$^{\circ}$ longitude region with associated water masers (note that sources without water masers also show a similar distribution).}
\label{fig:morph}
\end{center}
\end{figure}

Fig.~\ref{fig:col} presents a [3.6] - [4.5] versus [5.8] - [8.0] colour-colour plot using the \citeauthor{Gallaway13} determined IRAC magnitudes. The figure shows the majority of sources plotted on the corresponding figure 7 from \citet{Gallaway13} and thus is overall very similar. However, in our plot we have distinguished those 6.7-GHz methanol masers with and without both 12.2-GHz and 22-GHz counterparts. As expected, the maser-associated sources are much redder than the plotted comparison sources, and also often tend to have a relatively large excess of 4.5-$\mu$m flux \citep[as previously noted for methanol-associated sources][]{Breen12stats}. We find no significant differences in the distribution of the IRAC colours associated with and without either 12.2-GHz or water masers. Although we are dealing with only small numbers, a possible exception to this is a tendency for the water-associated sources to be located around the edges of the range of values occupied by the maser-associated sources in Fig.~\ref{fig:col}, with an absence of any sources falling in the centre of this range. In practical terms, this would mean that the water-associated sources tend to have either very similar magnitudes in adjacent IRAC bands, or far separated values; perhaps indicating a tendency for the water masers to be associated with EGO-like sources (sources with a 4.5-$\mu$m excess) or sources within regions of dominant PAH emission.

Several previous investigations have compared the GLIMPSE colours for sources harbouring different maser species, attempting to correlate maser evolutionary timelines with changes in the associated mid-infrared source properties. Both \citet{Ellingsen06} and \citet{Breen12stats} found that methanol maser sources also showing OH maser emission have relatively brighter 8-$\mu$m emission, the latter investigation also noting an increase in the 4.5-$\mu$m band for these sources. \citet{Ellingsen06} also found that class II methanol masers associated with class I sources generally have redder GLIMPSE point source colours, which they suggested was consistent with the class I methanol masers being associated with the younger sources.  Subsequent studies have shown that this is not the case for all class I methanol masers with several examples of their association with older sources, sometimes tracing expanding \ionhy regions \citep[e.g.][]{Voronkov06,Voronkov2010}. However, these previous results are small exceptions to an otherwise general lack of difference between the properties of GLIMPSE sources associated with different combinations of maser species (as seen here), leading to the suggestion that mid-infrared properties on this scale are less sensitive to small evolutionary changes than the different maser species. 

\citet{Gallaway13} inspected the infrared sources associated with each of the masers and characterised them as being either: 1) associated with an IRDC with no detectable IRAC counterpart; 2) associated with an IRAC source imbedded within an IRDC; 3) associated with an infrared-bright source, often extended and with no IRDC; 4) associated with neither an IRDC nor detectable emission counterpart. It was found that the majority of 6.7-GHz methanol masers (62 per cent) fall under the third category, followed by the second (21 per cent). We have investigated the distribution of \citeauthor{Gallaway13} categories for sources with and without 12.2-GHz methanol masers across the total longitude range (i.e. 186$^{\circ}$ to 20$^{\circ}$; plotted in Fig.~\ref{fig:morph}) and find that there is no difference between the distributions for 6.7-GHz masers with and without 12.2-GHz emission. We have also investigated the occurrence of water maser sources in the 6$^{\circ}$ to 20$^{\circ}$ longitude range and again find that there is no difference in the distribution of sources throughout the infrared morphology categories. 

We therefore corroborate a marked difference in the infrared properties of sources with and without masers \citep[as noted in previous large investigations;][]{Ellingsen06,Breen12stats,Gallaway13}, but there is little or no difference between sources with and without the different types of masers, either in colour or morphology.

%
%
%
%

\section{Summary}

We present a search for 12.2-GHz methanol masers towards a complete sample of MMB 6.7-GHz methanol masers detected in the Galactic longitude range 10$^{\circ}$ to 20$^{\circ}$. In total, 47 12.2-GHz methanol masers were detected towards 99 6.7-GHz targets. This catalogue is the third in this series, now covering all 6.7-GHz methanol masers across the longitude range 186$^{\circ}$ (through 0$^{\circ}$) to 20$^{\circ}$. The MMB survey extended further north to 60$^{\circ}$ and we will present 12.2-GHz observations towards these remaining sources following the publication of the 6.7-GHz methanol maser targets.  

We have compared the occurrence and characteristics of our 12.2-GHz methanol maser detections with: water maser emission observed towards all MMB sources in the 6$^{\circ}$ to 20$^{\circ}$ longitude range; 6.7-GHz methanol to OH maser peak flux density for all appropriate sources in the 186$^{\circ}$ (through 0$^{\circ}$) to 20$^{\circ}$ longitude range; and the association with GLIMPSE sources from \citet{Gallaway13}, a catalogue of sources not restricted to emission well fitted by a point sources (unlike many previous analyses). Each of these investigations primarily focused on furthering the evolutionary scenario for the different maser species. Through comparisons with the ratios of 6.7-GHz to OH maser peak flux density it appears that 12.2-GHz methanol masers experience a `turn over' in peak flux density, beginning the decline around the time that OH masers become detectable. It is likely that the proximity of the class II methanol and OH masers to the young stellar object makes them more direct and reliable tracers of physical changes in the nearby environment. We suggest that water masers do not neatly trace one evolutionary stage of high-mass star formation regions, rather that they accompany specific physical conditions (e.g. outflows and other shocks, PAH emission) which can occur at different times throughout the high-mass star formation lifetime.


\section*{Acknowledgments}

The Parkes telescope is part of the Australia Telescope which is funded by the Commonwealth of Australia for operation as a National Facility managed by CSIRO. Financial support for this work was provided by the Australian
Research Council. This research has made use of: NASA's Astrophysics
Data System Abstract Service; and  the SIMBAD data base, operated at CDS, Strasbourg,
France.

\end{document}